\documentclass[10pt,showpacs,onecolumn]{revtex4-1}
\usepackage{color}
\usepackage{graphicx}
\usepackage{amsmath}
\usepackage{amssymb}
\usepackage{epsfig}
\usepackage{epstopdf}
\usepackage{float}

\usepackage[absolute,showboxes]{textpos}  
\usepackage{}
\usepackage[
bookmarks=true,
colorlinks,
linkcolor=blue,
urlcolor=blue,
citecolor=blue,
plainpages=false,
pdfpagelabels,
final,
breaklinks=true
]{hyperref} 
\begin{document}

	\preprint{APS/123-QED}
        \title{Attosecond All-Optical Retrieval of Valley Polarization via Circular Dichroism in Transient Absorption}
	\author{Wenqing Li$^{1}$}
	\author{Xiaosong Zhu$^{1}$}\email{zhuxiaosong@hust.edu.cn}
        \author{Pengfei Lan$^{1,4}$}\email{pengfeilan@hust.edu.cn}
        \author{Kai Wang$^{1}$}
        \author{Wanzhu He$^{1}$}
        \author{Hannes H\"{u}bener$^{2}$}
        \author{Umberto De Giovannini$^{2,3}$}
	\author{Peixiang Lu$^{1,4}$}\email{lupeixiang@hust.edu.cn}

	\affiliation{%
		$^1$Wuhan National Laboratory for Optoelectronics and School of Physics,
		Huazhong University of Science and Technology, Wuhan 430074,
		China\\
            $^2$Max Planck Institute for the Structure and Dynamics of Matter and Center for Free-Electron Laser Science, Hamburg 22761, Germany\\
            $^3$Universit\`{a}  degli Studi di Palermo, Dipartimento di Fisica e Chimica-Emilio Segr\`{e}, Palermo I-90123, Italy\\
            $^4$Hubei Optical Fundamental Research Center, Wuhan 430074, China\\}

	
	\begin{abstract}

    We propose a scheme for retrieving the ultrafast valley polarization (VP) dynamics in two-dimensional hexagonal materials via attosecond circular dichroism (CD) transient absorption spectroscopy. This approach builds on the CD transition between the first and higher conduction bands induced by the circularly polarized probe pulses.
    The population imbalance at nonequivalent valleys in the first conduction band is proportionally mapped onto the difference in absorption coefficients of two probe pulses with opposite helicities, supporting an unprecedented quantitative retrieval of the corresponding VP dynamics with subfemtosecond time resolution.
    We theoretically demonstrate the scheme for h-BN and $\rm{MoS_2}$ through $ab\ initio$ calculations, achieving an accurate retrieval of the VP dynamics, particularly the transient VP switching processes, with a time resolution of 250 as. 		
	\end{abstract}
	
	\maketitle

   The electrons in two-dimensional (2D) hexagonal materials exhibit a valley degree of freedom in addition to charge and spin. This new degree of freedom has potential applications in next-generation electronics and optoelectronics \cite{Schaibley2016, Vitale2018}.  

    Selectively populating one valley, creating valley polarization (VP),  has been achieved by various methods \cite{Xiao2007, Xiao2012, OliaeiMotlagh2018, Silva2022, JimenezGalan2020, JimenezGalan2021, Mrudul2021, Sharma2022, Tyulnev2024, Mitra2024}, which is a key enabling feature of valleytronics. 
    Following the initialization, another crucial aspect is the detection of VP.   
    Over the past decade, some methods, such as the polarization-resolved photoluminescence spectroscopy and the Kerr rotation, have been used to detect the VP \cite{Zeng2012, Mak2012, Hsu2015, Wang2013, Kumar2014, Mai2014, Wang2013, Bertoni2016}. 
    Recently, approaches based on the second-harmonic spectroscopy \cite{JimenezGalan2021, Avetissian2023, Tyulnev2024} and high harmonic spectroscopy (HHS) \cite{JimenezGalan2020, JimenezGalan2021,  Mrudul2021, Mitra2024} have been proposed to detect VP in the femtosecond timescale. 

    Despite their utility, the methods above exhibit significant limitations. 
    Firstly, their time resolution is limited to the range of picoseconds to several tens of femtoseconds, which is not adequate to resolve the processes \textit{during} the switching of VP. 
    Besides, the time average of the signal over a long time scale \cite{Herrmann2023, Wang2018} or the lack of a clear and straightforward relationship between the observable quantities and the VP (such as in harmonics spectroscopies \cite{JimenezGalan2020, Mitra2024, Tyulnev2024}) hinders a precise and quantitative characterization of the time-dependent VP. 
    A recent study also raised concerns about the universal applicability of HHS in solids as a spectroscopic tool \cite{Neufeld2023}. 
    Overall, previous works usually employ the above techniques for investigations of the VP states \textit{after} a VP-switching operation or the valley relaxation process on the order of picoseconds.
   
    However, explicitly characterizing VP dynamics, particularly during the switching processes, provides key insights into the underlying physics governing the establishment and switching of VP. 
    This becomes increasingly important in light of the substantial research efforts currently being dedicated to the optical manipulation of VP, across various schemes and different regimes \cite{Xiao2012, OliaeiMotlagh2018, JimenezGalan2020, JimenezGalan2021, Mrudul2021, Silva2022, Sharma2022, Rana2023, Tyulnev2024, Mitra2024}, 
    and will thereby facilitate the applications for petahertz valleytronics. 
    These requirements highlight the urgency of approaches to accurately and even quantitatively capture the transient VP dynamics with subfemtosecond resolution. 
    However, as far as we know, the issue of reliable ultrafast readout of VP remains an open problem.

    Here, we propose and theoretically demonstrate an approach to retrieve the ultrafast VP dynamics via transient absorption spectroscopy \cite{Wu2016, Roerstad2017, Fazzi2013}, by leveraging the circular dichroism (CD) transition between the first conduction band (CB1) and higher unoccupied bands. 
     Benefiting from this configuration, 
     the population imbalance at nonequivalent valleys in CB1, namely the VP, is proportionally mapped on the difference between the absorption of two circularly polarized (CP) probe pulses with opposite helicities,  thereby supporting a quantitative retrieval of the  VP dynamics. 
    Besides, the large energy separation between the two levels supports subfemtosecond time resolution.
    To demonstrate our scheme, we perform numerical pump-probe experiments based on the time-dependent density functional theory (TDDFT) that simulate the measurements.
    We take prototypical 2D materials, h-BN and $\rm{MoS_2}$, as examples.
    Our results show that the VP dynamics, particularly the ultrafast switching, are faithfully retrieved with a $250\ \mathrm{as}$ resolution.

\begin{figure}[t]
	\includegraphics[width=14cm]{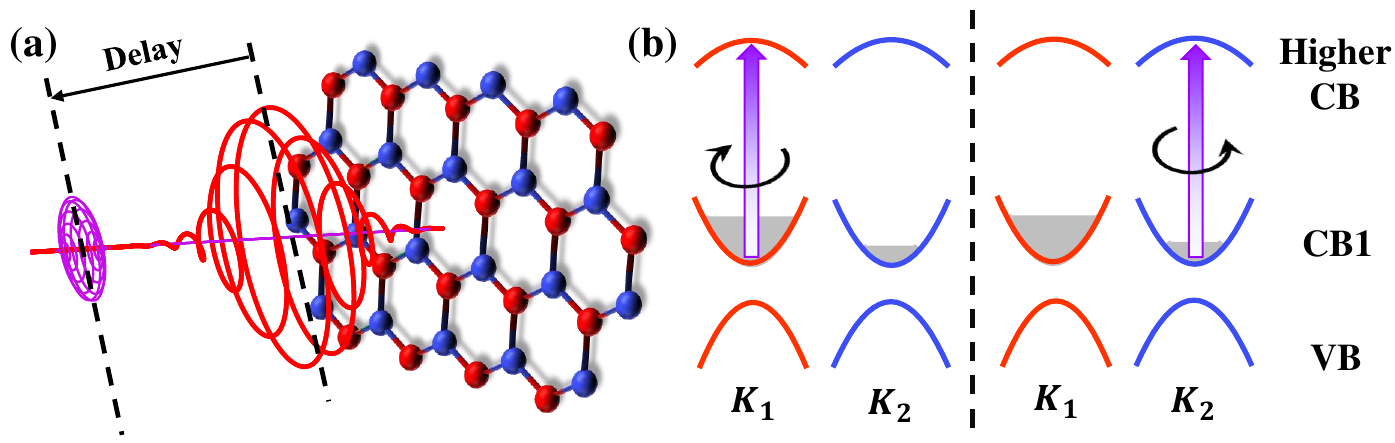}
	\caption{\label{fig1}%
			Schematic diagram of the scheme. 
   (a) The pump pulse (red line) manipulates the VP in 2D hexagonal material, which is subsequently probed by the CP probe pulse (purple line). 
   (b) Valley selective transition induced by probe pulses with opposite helicities.
   } 
	\end{figure}

      We first outline the fundamental idea in Fig.~\ref{fig1}.
      As shown in Fig.~\ref{fig1}(a), the VP is initialized by a pump pulse and subsequently probed by CP attosecond probe pulses with opposite helicities with adjustable delay. 
      For simplicity, we consider a CP pump pulse as an example. Note that the VP can also be achieved by various configurations, such as bi-circular fields \cite{JimenezGalan2020, Mrudul2021, Tyulnev2024, Mitra2024, Kfir2014, Zhu2022}. 
      Then, as depicted by the purple arrows in Fig.~\ref{fig1}(b), the probe pulses with different helicities will preferentially excite electrons around different valleys in CB1 to the higher unoccupied bands. 
      The valley-selective nature of the transition is intrinsically linked to the symmetry breaking of valley-active materials, underpinning the universal applicability of this approach \cite{Geondzhian2022, Feldman2024}. 
      Once the electron population imbalance at nonequivalent valleys in CB1 occurs, one probe pulse will excite more electrons, while the other pulse with the opposite helicity will excite fewer electrons.   
      Consequently, the absorbance of the two probe pulses will differ, resulting in remarkable CD signatures in the absorption spectra. 
      The CP attosecond probe pulses are experimentally feasible and have been applied to detect the ultrafast electron dynamics in diverse systems \cite{Kfir2014, Siegrist2019, Han2023, Geneaux2024}.

      In contrast to previous approaches involving the transition between VB and CB1, in our approach, the target band CB1 is the lower level and the upper levels are empty.
      Therefore, the VP and observed signals will exhibit a more direct relationship. 
      As derived in Sec.~\text{I} in Supplemental Material \cite{Supp},  the difference in the absorption coefficients for  $\pm$ probe pulses, denoted as $\Delta \mu = \mu_{+} - \mu_{-}$, where $+/–$ refers to the left or right circular polarization, 
      scales proportionally to the population imbalance in CB1:
    \begin{equation}\label{linear}
		 \Delta \mu = \mu_{+} - \mu_{-} \propto N^{\mathrm{\mathbf{K_1}}}_{\rm{CB1}} - N^{\mathrm{\mathbf{K_2}}}_{\rm CB1}.
    \end{equation}
    $N^{\mathrm{\mathbf{K_1}}}_{\rm{CB1}}$($N^{\mathrm{\mathbf{K_2}}}_{\rm CB1}$) is the number of electrons  around 
    $\mathrm{\mathbf{K_1}}$($\mathrm{\mathbf{K_2}}$) valley in CB1. 
    Namely, the observable $\Delta\mu$ scales proportionally with the VP and hence is a direct indicator for the magnitude of this effect. 

      Additionally, the upper levels in our approach can provide a larger energy separation up to the ultraviolet range,
      which supports the subfemtosecond time resolution.
      By applying attosecond probe pulses and scanning the pump-probe delay, the VP dynamics can be retrieved by the resultant CD in the transient absorption spectra with attosecond resolution.  

     Firstly, we demonstrate a numerical pump-probe experiment in h-BN to validate the above expectations.
      The VP is prepared by applying a CP pump pulse with vector potential:  
      \begin{equation}\label{pump1}
	\begin{aligned}
        A_{\rm{pump}} = \frac{F_1}{\omega_1}f(t,\tau) [\sin(\omega_1(t - \tau))\hat{\mathbf{x}}  -\cos(\omega_1(t - \tau))\hat{\mathbf{y}}].
	\end{aligned}
       \end{equation}
     $f(t,\tau)$ represents a Gaussian envelope centered at $\tau$ with the full width at half maximum (FWHM) of $ 1.25\ {\rm fs}$. Here, $\tau = 5\ {\rm fs} $, 
     $F_1 = 0.1\ {\rm V/\AA}$, and $\omega_1 = 4.6 \ {\rm eV/\hbar}$ corresponding to the resonant excitation from VB to CB1 around the 
     valleys in h-BN.  The electron dynamics in 2D materials is described by  TDDFT. The simulations are implemented with Octopus \cite{ Marques2003, Castro2006, Andrade2015, TancogneDejean2020a}. Technical details are delegated in Sec.~\text{II} in Supplemental Material \cite{Supp}. 

     In the simulation, we can directly quantify the VP by counting excited electrons in CB1, i.e. calculating $N^{\mathrm{\mathbf{K_1}}}_{\rm{CB1}}$ ($N^{\mathrm{\mathbf{K_2}}}_{\rm CB1}$), via integration of the momentum-resolved electron population in CB1 inside a circle of radius $\mathcal{R}$ around $\mathbf{K_1}$ ($\mathbf{K_2}$) in the k-space. Here, $\mathcal{R}$ takes one-third of the distance between two neighboring valleys.  
     Due to the optical valley selection rules \cite{Xiao2007, Xiao2012}, the electrons around the $\mathrm{\mathbf{K_1}}$ valley are selectively excited to CB1 (c.f. Fig.~\ref{fig3}(b)). In this case, $\Delta N_{\rm{CB1}} = N^{\mathrm{\mathbf{K_1}}}_{\rm{CB1}} - N^{\mathrm{\mathbf{K_2}}}_{\rm CB1} > 0$.

     Subsequently, the prepared VP  is probed by CP attosecond pulses: 
     \begin{equation}\label{probe}
	\begin{aligned}
        A^{\pm}_{\rm{probe}} = \frac{F_2}{\omega_2}f_p(t, \tau_p)[\cos(\omega_2(t - \tau_p))\hat{\mathbf{x}} \mp \sin(\omega_2(t - \tau_p))\hat{\mathbf{y}}].
	\end{aligned}
      \end{equation}
     $f_p(t, \tau_p) = \sin^2(\frac{\pi (t - \tau_p)}{T_{\rm{probe}}})$ represents the envelope centered at $\tau_p$ with the full duration $T_{\rm{probe}} =  1\ {\rm fs}$. 
     $F_2 = 0.007\ {\rm V/\AA}$.
     $\omega_2 = 7.5 \ {\rm eV/\hbar}$ matches the bandgap between CB1 and the second conduction band (CB2) at the valleys.
     In this case, the $\pm$ probe pulses will preferentially excite electrons around respective valleys in CB1 to CB2.

     The valley-selectivity of the transition within the conduction bands is confirmed by calculating the squared modulus of corresponding momentum matrix elements for circular polarization of opposite helicities $ | \langle u_{\rm{CB1}}| \hat P_{\pm} |u_{\rm{CB2}}\rangle |^2$. 
     Here, $\hat P_{\pm} = \hat P_{x} \pm i\hat P_{y}$ with $x/y$ component of the momentum operator $\hat P_{x/y}$ and $|u_{\rm{CB1,2}}\rangle$ represents the periodic part of the Bloch function of CB1 and CB2 \cite{Berkelbach2015, Li2019, Li2023}, which is obtained from the ground-state DFT calculation. 
     The result indicates that, for a prepared VP state, different numbers of electrons will be excited into CB2 by the  $\pm$ probe pulses.

             \begin{figure}[t]
		\includegraphics[width=10cm]{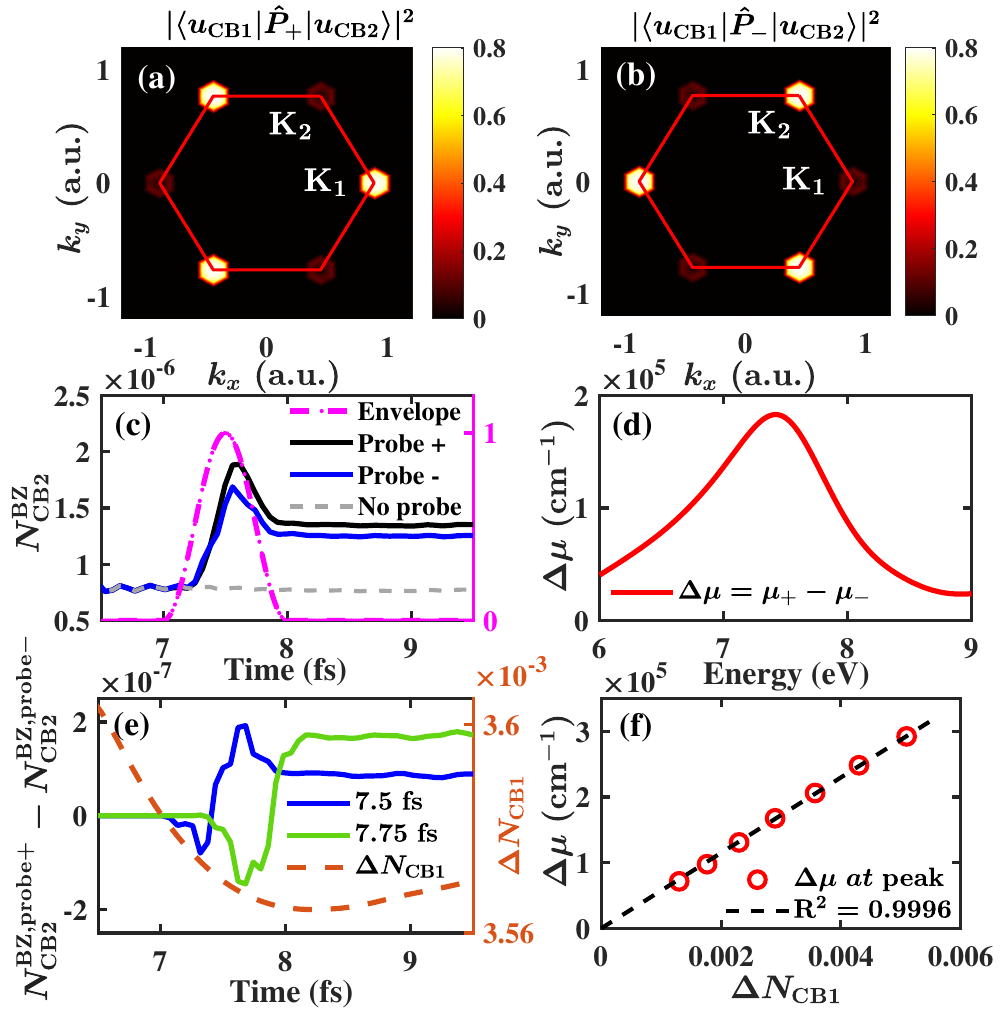}
		\caption{\label{fig2}
        (a) (b) The squared modulus of momentum matrix elements  $ | \langle u_{\rm{CB1}}| \hat P_{\pm} |u_{\rm{CB2}}\rangle |^2$ in h-BN, for the circular polarization with  $+$ and $-$ helicity, respectively. 
        (c)	 Time evolution of the electron population in CB2 for   $+$ (solid black line) and $-$ (solid blue line) probe pulses, and without probe pulse (dashed grey line).  
        The dashed violet line represents the envelope of probe pulses centered at $7.5\  \mathrm{fs}$.
        (d) The difference in the absorption coefficients between the probe pulses with opposite helicities ($\Delta \mu = \mu_{+} - \mu_{-}$). 
        (e) Time evolution of $N_{\rm{CB2}}^{\rm{BZ, probe +}} - N_{\rm{CB2}}^{\rm{BZ, probe -}}$ for probe pulses centered at $7.5\  \mathrm{fs}$ (solid blue line) and $7.75\  \mathrm{fs}$ (solid green line). The dashed line represents $\Delta N_{\rm{CB1}}$, characterizing the VP.   
        (f) $\Delta \mu$ at $7.5\ \mathrm{eV}$ vs. $\Delta N_{\rm{CB1}}$ (red circles) compared to discussion in Eq.~(\ref{linear}).} 
	\end{figure}
     
     Taking probe pulses centered at $\tau_p = 7.5\ \rm{fs}$ as examples, we calculate the time evolution of the electron population in CB2  ($N_{\rm CB2}^{\rm{BZ}}(t)$, with superscript BZ representing integration over the Brillouin zone), as shown in Fig.~\ref{fig2}(c). Momentum-resolved population changes induced by the probe pulses are given in Sec. III in Supplemental Material \cite{Supp}. 
     Comparing the black curve $N_{\rm{CB2}}^{\rm{BZ, probe +}}$ with the blue curve $N_{\rm{CB2}}^{\rm{BZ, probe -}}$ in Fig.~\ref{fig2}(c), more electrons are excited by the $+$ probe pulse than the $-$ pulse,
     implying that the system absorbs more photons under the $+$ probe pulse. 
     This is validated in Fig.~\ref{fig2}(d), where we calculate the difference of the energy-resolved transient absorption coefficients for the $\pm$ probe pulses $\Delta \mu (\omega,\tau_p = 7.5\ \rm{fs}) = \mu_{+} - \mu_{-}$. 
     Details of the calculations for $\mu_{\pm}$ are provided in Sec.~\text{IV} in Supplemental Material \cite{Supp}. 
     One can see that a significant CD signal is obtained, 
     reflecting the prepared VP. The CD signal reaches its maximum around  $7.5\ \rm{eV}$, which corresponds to the energy difference between CB1 and CB2 at the valleys. 
     The solid blue and green lines in Fig.~\ref{fig2}(e) show 
     $N_{\rm{CB2}}^{\rm{BZ, probe +}}(t) - N_{\rm{CB2}}^{\rm{BZ, probe -}}(t)$ for 
     pulses with different probe times $ \tau _p = 7.5\  \mathrm{fs}$ and $7.75\  \mathrm{fs}$, respectively.
     The clear separation between the two curves shows that distinct electron dynamics occur under the probe pulses at different times, leading to different net photon absorption. The distinct responses indicate that the probe can distinguish different transient VP states separated by 250 $\rm{as}$, underscoring the capability for subfemtosecond temporal resolution. 
     
     To examine the correspondence between the VP ($\Delta N_{\rm{CB1}}$)  and the observable CD in the absorption coefficient ($\Delta \mu$), we prepare VP states with different $\Delta N_{\rm{CB1}}$ by varying $F_1$ of the pump pulse, and then calculate $\Delta \mu$ of these states for the probe pulses fixed at $ \tau _p = 7.5\  \mathrm{fs}$.   
     The resulting $\Delta \mu$ at $7.5 \ {\rm eV}$ 
     as a function of $\Delta N_{\rm{CB1}}$ are summarized in Fig.~\ref{fig2}(f) by the red circles. 
     As the dashed linearly fitting line indicates, $\Delta \mu$ is proportional to $\Delta N_{\rm{CB1}}$, which is consistent with Eq.~(\ref{linear}), exhibiting a direct and simple one-to-one mapping. 
     It enables a precise retrieval of time-dependent VP dynamics.

         \begin{figure}[htbp]

	\includegraphics[width=10cm]{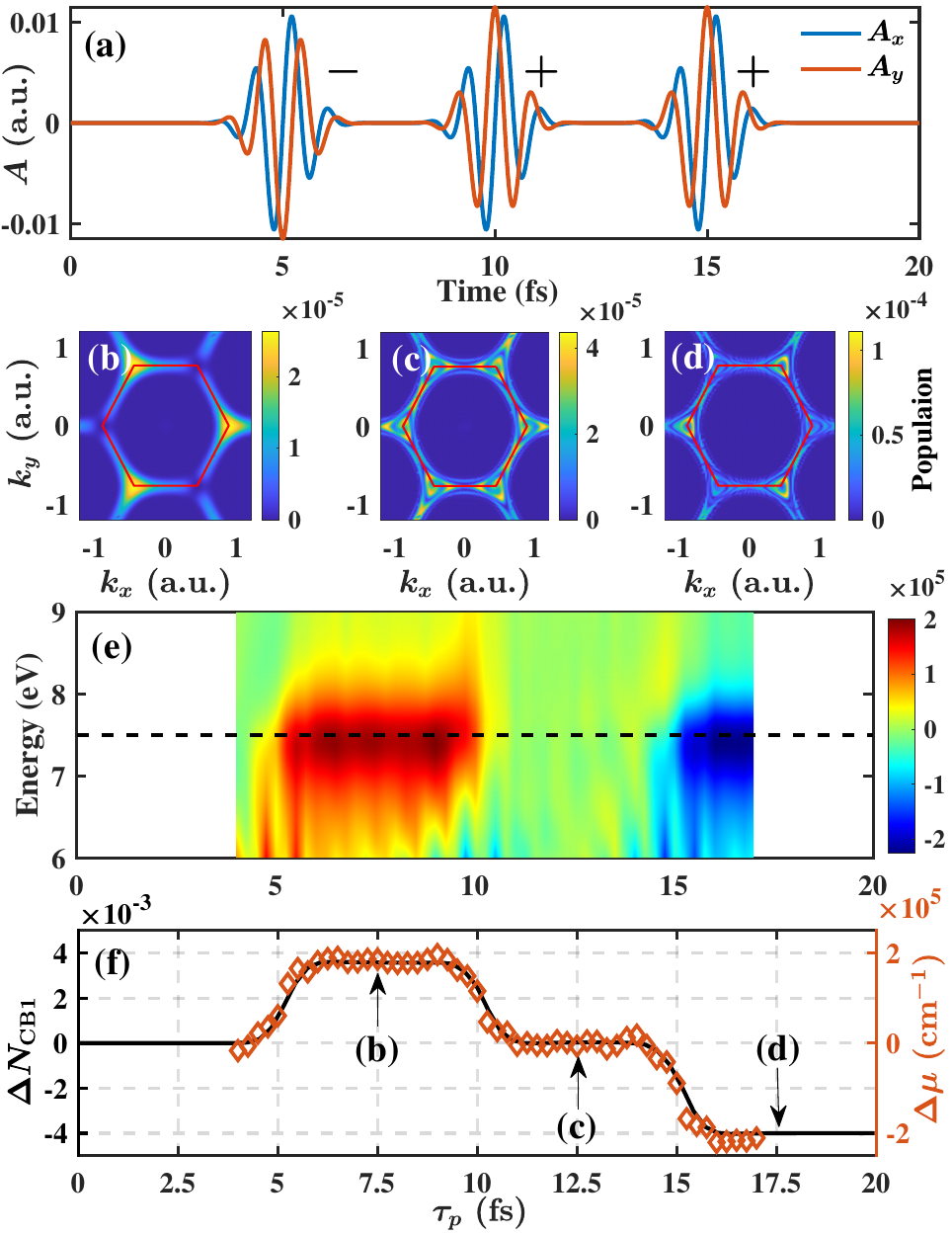}
	\caption{\label{fig3}
      Retrieval of the ultrafast VP dynamics in h-BN. (a) The vector potential of the switching pulses given in atomic units (a.u.). 
      The momentum-resolved electron populations in CB1 at  (b) $7.5\ \mathrm{fs}$,  (c) $12.5\ \mathrm{fs}$, and  (d) $17.5\ \mathrm{fs}$. 
      (e) The calculated  $\Delta \mu(\omega,\tau_p)$.   
      (f)   Time evolution of $\Delta N_{\rm{CB1}}$ (solid black line) and $\Delta \mu(\omega,\tau_p)$ (orange diamonds) at $7.5\ {\rm eV}$ indicated by the dashed black line in (e). 
         } 
	\end{figure}

      As an illustration, we demonstrate our scheme to retrieve the VP dynamics in h-BN driven by three switching pulses. The three switching pulses are the same as the pump pulse used for Fig.~\ref{fig2} but with alternating helicities $-$, $+$, $+$, and they are centered at $\tau =  5\ {\rm fs} $, $10\ {\rm fs} $, $15\ {\rm fs}$, as shown in Fig.~\ref{fig3}(a). 
     All of the protocol is performed on timescales one order of magnitude shorter than valley lifetimes \cite{Lively2024}.

    The switching pulses will selectively and sequentially excite the electrons around the $\mathrm{\mathbf{K_1}}$,  $\mathrm{\mathbf{K_2}}$, and $\mathrm{\mathbf{K_2}}$ valleys.       
    The momentum-resolved electron populations in CB1 at $7.5\ \mathrm{fs}$, $12.5\ \mathrm{fs}$, and $17.5\ \mathrm{fs}$ are shown in Figs.~\ref{fig3}(b)-(d), respectively. Meanwhile, the time-dependent $\Delta N_{\rm{CB1}}$ is shown in  Fig.~\ref{fig3}(f). 
    The results show that the obtained VP is turned on, off, and reversed. 

    To probe the VP dynamics, the probe pulses described by Eq.~(\ref{probe}) at various $\tau_p$ ($4\ {\rm fs}< \tau_p < 17\ {\rm fs}$) with time step of 250 as are applied. 
    The $\Delta \mu (\omega, \tau_p)$, namely the CD in the absorption spectra, is calculated and displayed in Fig.~\ref{fig3}(e). Meanwhile, a slice along 7.5 eV in Fig.~\ref{fig3}(e) is depicted as the orange diamonds in Fig.~\ref{fig3}(f).
    Remarkably, one can see that the magnitude of $\Delta \mu$ faithfully reproduces the time variation of the VP. This follows from the linear correspondence between the observable quantity and VP as discussed with Eq.~(\ref{linear}) and verified in Fig.~\ref{fig2}(f). 
    Note that, the proportionality constant in Eq.~(\ref{linear}) can be experimentally determined by one auxiliary measurement of an absolute $\Delta \rm{N_{CB1}}$ 
    (e.g., with ARPES \cite{Bertoni2016}) in the static or slowly varying VP condition.
    This means that here we present a scheme for the precise measurement of the absolute value of the ultrafast VP.  
    Importantly, our method not only detects the switched VP states after each pump pulse. The sequential switching in the vicinity of $5\ \mathrm{fs}$, $10\ \mathrm{fs}$, and $15\ \mathrm{fs}$ are clearly resolved within 250 attosecond. Such a time resolution is unprecedented. 
    
    
    
    %

\begin{figure}[htbp]

	\includegraphics[width=10cm]{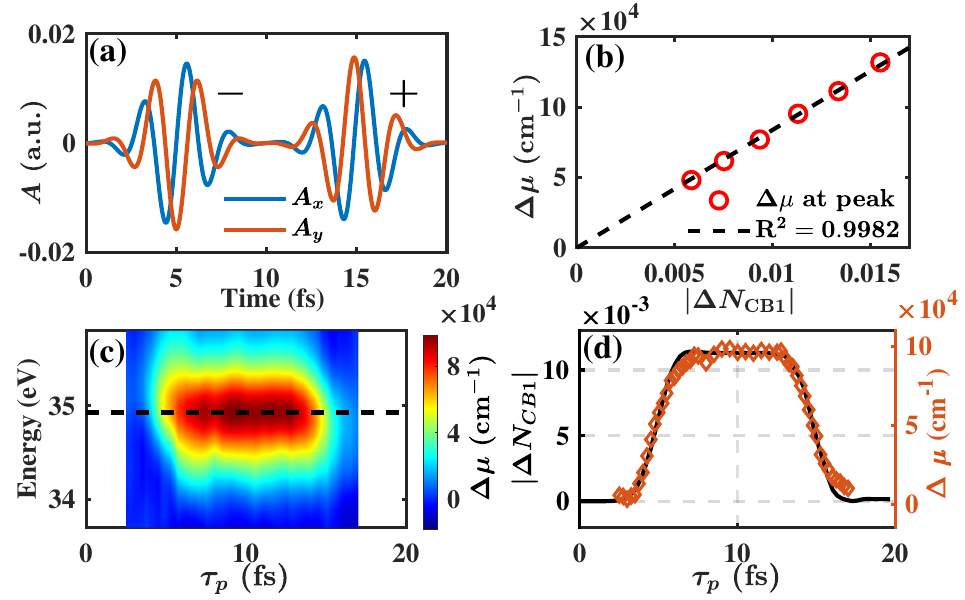}
	\caption{\label{fig4}
      Retrieval of the ultrafast VP dynamics in MoS$_2$. (a) The vector potential of the switching pulses.
      (b) $\Delta \mu$ at $34.9\ {\rm eV}$ vs. $|\Delta N_{\rm{CB1}}|$ (red circles) compared to discussion in Eq.~(\ref{linear}), where $34.9\ {\rm eV}$ is the peak of the CD absorption spectra.  
      (c) The calculated  $\Delta \mu(\omega,\tau_p)$.  
      (d) 
      The time evolution of $|\Delta N_{\rm{CB1}}|$ (solid black line) and $\Delta \mu(\omega,\tau_p)$ (orange diamonds) at $34.9\ {\rm eV}$ indicated by the dashed black line in (c).
        } 
	\end{figure}

    We next explore the versatility of this scheme by applying it to MoS$_2$, another promising material for valleytronics. 
    The time-dependent VP is manipulated by two CP pulses (given by Eq.~(\ref{pump1})) with $-$ and $+$ helicities as shown in Fig.~\ref{fig4}(a). The pulse parameters are: $F_1 = 0.05\ {\rm V/\AA}$ and $\omega_1 = 1.68 \ {\rm eV/\hbar}$ to match the bandgap between VB and CB1 at the valleys in MoS$_2$. The FWHM of the Gaussian envelope is $3.5\ \rm{fs}$, and the two pulses are centered at $\tau = 5\ {\rm fs}$ and $10\ {\rm fs}$, respectively. 
    The time-dependent $|\Delta N_{\rm{CB1}}|$ is illustrated with the solid black line in Fig.~\ref{fig4}(d), showing that the prepared VP is turned on and then turned off.  

    The parameters of probe pulses are the same as those in h-BN except for  $\omega_2 = 35 \ {\rm eV/\hbar}$. 
    Following the procedure in Fig.~\ref{fig2}(f),  we examine the correspondence between the VP and the observable CD signals in MoS$_2$ as shown in Fig.~\ref{fig4}(b), confirming the linear correspondence from Eq.~(\ref{linear}). 
    The delay-dependent $\Delta \mu (\omega, \tau_p)$ is calculated and displayed in Fig.~\ref{fig4}(c), and the slice along $34.9\ {\rm eV}$ (dashed black line)
    is depicted as the orange diamonds in Fig.~\ref{fig4}(d). 
    One can see that the magnitude of $\Delta \mu$ 
    faithfully reproduces the VP dynamics in MoS$_2$ with a $250\ \mathrm{as}$ time resolution.

In summary, we propose an all-optical approach that supports an unprecedented quantitative retrieval of the VP dynamics with attosecond time resolution. This is achieved through transient absorption spectroscopy leveraging the CD transition between CB1 and higher unoccupied bands. In this scenario, the VP is proportionally mapped to the observable, enabling the quantitative measurement. Meanwhile, the larger energy separation between CB1 and higher bands supports the sub-femtosecond time resolution. 
We numerically demonstrate this approach in h-BN and MoS$_2$. The results show that the ultrafast VP dynamics, particularly the transient VP switching processes, are accurately retrieved with a $250\ \mathrm{as}$ time resolution.

	\section*{Acknowledgement}
	This work was supported by National Key Research and Development Program under grant No. 2023YFA1406800, the National Natural Science Foundation of China under grants Nos. 12174134,  12450406,  12225406, and  12021004,  European Union – Next Generation EU through project THENCE – Partenariato Esteso NQSTI – PE00000023 – Spoke 2, and the Marie Sklodowska-Curie Doctoral Networks TIMES grant No. 101118915 and SPARKLE grant No. 101169225.  
      The computation is completed in the HPC Platform of Huazhong University of Science and Technology.
	\bibliography{liwenqing}

\begin{thebibliography}{51}%
\makeatletter
\providecommand \@ifxundefined [1]{%
 \@ifx{#1\undefined}
}%
\providecommand \@ifnum [1]{%
 \ifnum #1\expandafter \@firstoftwo
 \else \expandafter \@secondoftwo
 \fi
}%
\providecommand \@ifx [1]{%
 \ifx #1\expandafter \@firstoftwo
 \else \expandafter \@secondoftwo
 \fi
}%
\providecommand \natexlab [1]{#1}%
\providecommand \enquote  [1]{``#1''}%
\providecommand \bibnamefont  [1]{#1}%
\providecommand \bibfnamefont [1]{#1}%
\providecommand \citenamefont [1]{#1}%
\providecommand \href@noop [0]{\@secondoftwo}%
\providecommand \href [0]{\begingroup \@sanitize@url \@href}%
\providecommand \@href[1]{\@@startlink{#1}\@@href}%
\providecommand \@@href[1]{\endgroup#1\@@endlink}%
\providecommand \@sanitize@url [0]{\catcode `\\12\catcode `\$12\catcode `\&12\catcode `\#12\catcode `\^12\catcode `\_12\catcode `\%12\relax}%
\providecommand \@@startlink[1]{}%
\providecommand \@@endlink[0]{}%
\providecommand \url  [0]{\begingroup\@sanitize@url \@url }%
\providecommand \@url [1]{\endgroup\@href {#1}{\urlprefix }}%
\providecommand \urlprefix  [0]{URL }%
\providecommand \Eprint [0]{\href }%
\providecommand \doibase [0]{http://dx.doi.org/}%
\providecommand \selectlanguage [0]{\@gobble}%
\providecommand \bibinfo  [0]{\@secondoftwo}%
\providecommand \bibfield  [0]{\@secondoftwo}%
\providecommand \translation [1]{[#1]}%
\providecommand \BibitemOpen [0]{}%
\providecommand \bibitemStop [0]{}%
\providecommand \bibitemNoStop [0]{.\EOS\space}%
\providecommand \EOS [0]{\spacefactor3000\relax}%
\providecommand \BibitemShut  [1]{\csname bibitem#1\endcsname}%
\let\auto@bib@innerbib\@empty
\bibitem [{\citenamefont {Schaibley}\ \emph {et~al.}(2016)\citenamefont {Schaibley}, \citenamefont {Yu}, \citenamefont {Clark}, \citenamefont {Rivera}, \citenamefont {Ross}, \citenamefont {Seyler}, \citenamefont {Yao},\ and\ \citenamefont {Xu}}]{Schaibley2016}%
  \BibitemOpen
  \bibfield  {author} {\bibinfo {author} {\bibfnamefont {J.~R.}\ \bibnamefont {Schaibley}}, \bibinfo {author} {\bibfnamefont {H.}~\bibnamefont {Yu}}, \bibinfo {author} {\bibfnamefont {G.}~\bibnamefont {Clark}}, \bibinfo {author} {\bibfnamefont {P.}~\bibnamefont {Rivera}}, \bibinfo {author} {\bibfnamefont {J.~S.}\ \bibnamefont {Ross}}, \bibinfo {author} {\bibfnamefont {K.~L.}\ \bibnamefont {Seyler}}, \bibinfo {author} {\bibfnamefont {W.}~\bibnamefont {Yao}}, \ and\ \bibinfo {author} {\bibfnamefont {X.}~\bibnamefont {Xu}},\ }\href {https://doi.org/10.1038/natrevmats.2016.55} {\bibfield  {journal} {\bibinfo  {journal} {Nat. Rev. Mater.}\ }\textbf {\bibinfo {volume} {1}},\ \bibinfo {pages} {16055} (\bibinfo {year} {2016})}\BibitemShut {NoStop}%
\bibitem [{\citenamefont {Vitale}\ \emph {et~al.}(2018)\citenamefont {Vitale}, \citenamefont {Nezich}, \citenamefont {Varghese}, \citenamefont {Kim}, \citenamefont {Gedik}, \citenamefont {Jarillo‐Herrero}, \citenamefont {Xiao},\ and\ \citenamefont {Rothschild}}]{Vitale2018}%
  \BibitemOpen
  \bibfield  {author} {\bibinfo {author} {\bibfnamefont {S.~A.}\ \bibnamefont {Vitale}}, \bibinfo {author} {\bibfnamefont {D.}~\bibnamefont {Nezich}}, \bibinfo {author} {\bibfnamefont {J.~O.}\ \bibnamefont {Varghese}}, \bibinfo {author} {\bibfnamefont {P.}~\bibnamefont {Kim}}, \bibinfo {author} {\bibfnamefont {N.}~\bibnamefont {Gedik}}, \bibinfo {author} {\bibfnamefont {P.}~\bibnamefont {Jarillo‐Herrero}}, \bibinfo {author} {\bibfnamefont {D.}~\bibnamefont {Xiao}}, \ and\ \bibinfo {author} {\bibfnamefont {M.}~\bibnamefont {Rothschild}},\ }\href {https://doi.org/10.1002/smll.201801483} {\bibfield  {journal} {\bibinfo  {journal} {Small}\ }\textbf {\bibinfo {volume} {14}},\ \bibinfo {pages} {1801483} (\bibinfo {year} {2018})}\BibitemShut {NoStop}%
\bibitem [{\citenamefont {Xiao}\ \emph {et~al.}(2007)\citenamefont {Xiao}, \citenamefont {Yao},\ and\ \citenamefont {Niu}}]{Xiao2007}%
  \BibitemOpen
  \bibfield  {author} {\bibinfo {author} {\bibfnamefont {D.}~\bibnamefont {Xiao}}, \bibinfo {author} {\bibfnamefont {W.}~\bibnamefont {Yao}}, \ and\ \bibinfo {author} {\bibfnamefont {Q.}~\bibnamefont {Niu}},\ }\href {\doibase 10.1103/physrevlett.99.236809} {\bibfield  {journal} {\bibinfo  {journal} {Phys. Rev. Lett.}\ }\textbf {\bibinfo {volume} {99}},\ \bibinfo {pages} {236809} (\bibinfo {year} {2007})}\BibitemShut {NoStop}%
\bibitem [{\citenamefont {Xiao}\ \emph {et~al.}(2012)\citenamefont {Xiao}, \citenamefont {Liu}, \citenamefont {Feng}, \citenamefont {Xu},\ and\ \citenamefont {Yao}}]{Xiao2012}%
  \BibitemOpen
  \bibfield  {author} {\bibinfo {author} {\bibfnamefont {D.}~\bibnamefont {Xiao}}, \bibinfo {author} {\bibfnamefont {G.-B.}\ \bibnamefont {Liu}}, \bibinfo {author} {\bibfnamefont {W.}~\bibnamefont {Feng}}, \bibinfo {author} {\bibfnamefont {X.}~\bibnamefont {Xu}}, \ and\ \bibinfo {author} {\bibfnamefont {W.}~\bibnamefont {Yao}},\ }\href {\doibase 10.1103/physrevlett.108.196802} {\bibfield  {journal} {\bibinfo  {journal} {Phys. Rev. Lett.}\ }\textbf {\bibinfo {volume} {108}},\ \bibinfo {pages} {196802} (\bibinfo {year} {2012})}\BibitemShut {NoStop}%
\bibitem [{\citenamefont {Oliaei~Motlagh}\ \emph {et~al.}(2018)\citenamefont {Oliaei~Motlagh}, \citenamefont {Wu}, \citenamefont {Apalkov},\ and\ \citenamefont {Stockman}}]{OliaeiMotlagh2018}%
  \BibitemOpen
  \bibfield  {author} {\bibinfo {author} {\bibfnamefont {S.~A.}\ \bibnamefont {Oliaei~Motlagh}}, \bibinfo {author} {\bibfnamefont {J.-S.}\ \bibnamefont {Wu}}, \bibinfo {author} {\bibfnamefont {V.}~\bibnamefont {Apalkov}}, \ and\ \bibinfo {author} {\bibfnamefont {M.~I.}\ \bibnamefont {Stockman}},\ }\href {\doibase 10.1103/physrevb.98.081406} {\bibfield  {journal} {\bibinfo  {journal} {Phys. Rev. B}\ }\textbf {\bibinfo {volume} {98}},\ \bibinfo {pages} {081406} (\bibinfo {year} {2018})}\BibitemShut {NoStop}%
\bibitem [{\citenamefont {Silva}\ \emph {et~al.}(2022)\citenamefont {Silva}, \citenamefont {Ivanov},\ and\ \citenamefont {Jim{\'e}nez-Gal{\'a}n}}]{Silva2022}%
  \BibitemOpen
  \bibfield  {author} {\bibinfo {author} {\bibfnamefont {R.~E.~F.}\ \bibnamefont {Silva}}, \bibinfo {author} {\bibfnamefont {M.}~\bibnamefont {Ivanov}}, \ and\ \bibinfo {author} {\bibfnamefont {{\'A}.}~\bibnamefont {Jim{\'e}nez-Gal{\'a}n}},\ }\href {\doibase 10.1364/oe.460291} {\bibfield  {journal} {\bibinfo  {journal} {Opt. Express}\ }\textbf {\bibinfo {volume} {30}},\ \bibinfo {pages} {30347} (\bibinfo {year} {2022})}\BibitemShut {NoStop}%
\bibitem [{\citenamefont {Jim{\'e}nez-Gal{\'a}n}\ \emph {et~al.}(2020)\citenamefont {Jim{\'e}nez-Gal{\'a}n}, \citenamefont {Silva}, \citenamefont {Smirnova},\ and\ \citenamefont {Ivanov}}]{JimenezGalan2020}%
  \BibitemOpen
  \bibfield  {author} {\bibinfo {author} {\bibfnamefont {{\'A}.}~\bibnamefont {Jim{\'e}nez-Gal{\'a}n}}, \bibinfo {author} {\bibfnamefont {R.~E.~F.}\ \bibnamefont {Silva}}, \bibinfo {author} {\bibfnamefont {O.}~\bibnamefont {Smirnova}}, \ and\ \bibinfo {author} {\bibfnamefont {M.}~\bibnamefont {Ivanov}},\ }\href {\doibase 10.1038/s41566-020-00717-3} {\bibfield  {journal} {\bibinfo  {journal} {Nat. Photon.}\ }\textbf {\bibinfo {volume} {14}},\ \bibinfo {pages} {728} (\bibinfo {year} {2020})}\BibitemShut {NoStop}%
\bibitem [{\citenamefont {Jim{\'e}nez-Gal{\'a}n}\ \emph {et~al.}(2021)\citenamefont {Jim{\'e}nez-Gal{\'a}n}, \citenamefont {Silva}, \citenamefont {Smirnova},\ and\ \citenamefont {Ivanov}}]{JimenezGalan2021}%
  \BibitemOpen
  \bibfield  {author} {\bibinfo {author} {\bibfnamefont {{\'A}.}~\bibnamefont {Jim{\'e}nez-Gal{\'a}n}}, \bibinfo {author} {\bibfnamefont {R.~E.~F.}\ \bibnamefont {Silva}}, \bibinfo {author} {\bibfnamefont {O.}~\bibnamefont {Smirnova}}, \ and\ \bibinfo {author} {\bibfnamefont {M.}~\bibnamefont {Ivanov}},\ }\href {\doibase 10.1364/optica.404257} {\bibfield  {journal} {\bibinfo  {journal} {Optica}\ }\textbf {\bibinfo {volume} {8}},\ \bibinfo {pages} {277} (\bibinfo {year} {2021})}\BibitemShut {NoStop}%
\bibitem [{\citenamefont {Mrudul}\ \emph {et~al.}(2021)\citenamefont {Mrudul}, \citenamefont {Jim{\'e}nez-Gal{\'a}n}, \citenamefont {Ivanov},\ and\ \citenamefont {Dixit}}]{Mrudul2021}%
  \BibitemOpen
  \bibfield  {author} {\bibinfo {author} {\bibfnamefont {M.~S.}\ \bibnamefont {Mrudul}}, \bibinfo {author} {\bibfnamefont {{\'A}.}~\bibnamefont {Jim{\'e}nez-Gal{\'a}n}}, \bibinfo {author} {\bibfnamefont {M.}~\bibnamefont {Ivanov}}, \ and\ \bibinfo {author} {\bibfnamefont {G.}~\bibnamefont {Dixit}},\ }\href {\doibase 10.1364/optica.418152} {\bibfield  {journal} {\bibinfo  {journal} {Optica}\ }\textbf {\bibinfo {volume} {8}},\ \bibinfo {pages} {422} (\bibinfo {year} {2021})}\BibitemShut {NoStop}%
\bibitem [{\citenamefont {Sharma}\ \emph {et~al.}(2022)\citenamefont {Sharma}, \citenamefont {Elliott},\ and\ \citenamefont {Shallcross}}]{Sharma2022}%
  \BibitemOpen
  \bibfield  {author} {\bibinfo {author} {\bibfnamefont {S.}~\bibnamefont {Sharma}}, \bibinfo {author} {\bibfnamefont {P.}~\bibnamefont {Elliott}}, \ and\ \bibinfo {author} {\bibfnamefont {S.}~\bibnamefont {Shallcross}},\ }\href {\doibase 10.1364/optica.458991} {\bibfield  {journal} {\bibinfo  {journal} {Optica}\ }\textbf {\bibinfo {volume} {9}},\ \bibinfo {pages} {947} (\bibinfo {year} {2022})}\BibitemShut {NoStop}%
\bibitem [{\citenamefont {Tyulnev}\ \emph {et~al.}(2024)\citenamefont {Tyulnev}, \citenamefont {Jim{\'e}nez-Gal{\'a}n}, \citenamefont {Poborska}, \citenamefont {Vamos}, \citenamefont {Russell}, \citenamefont {Tani}, \citenamefont {Smirnova}, \citenamefont {Ivanov}, \citenamefont {Silva},\ and\ \citenamefont {Biegert}}]{Tyulnev2024}%
  \BibitemOpen
  \bibfield  {author} {\bibinfo {author} {\bibfnamefont {I.}~\bibnamefont {Tyulnev}}, \bibinfo {author} {\bibfnamefont {{\'A}.}~\bibnamefont {Jim{\'e}nez-Gal{\'a}n}}, \bibinfo {author} {\bibfnamefont {J.}~\bibnamefont {Poborska}}, \bibinfo {author} {\bibfnamefont {L.}~\bibnamefont {Vamos}}, \bibinfo {author} {\bibfnamefont {P.~S.~J.}\ \bibnamefont {Russell}}, \bibinfo {author} {\bibfnamefont {F.}~\bibnamefont {Tani}}, \bibinfo {author} {\bibfnamefont {O.}~\bibnamefont {Smirnova}}, \bibinfo {author} {\bibfnamefont {M.}~\bibnamefont {Ivanov}}, \bibinfo {author} {\bibfnamefont {R.~E.~F.}\ \bibnamefont {Silva}}, \ and\ \bibinfo {author} {\bibfnamefont {J.}~\bibnamefont {Biegert}},\ }\href {\doibase 10.1038/s41586-024-07156-y} {\bibfield  {journal} {\bibinfo  {journal} {Nature}\ }\textbf {\bibinfo {volume} {628}},\ \bibinfo {pages} {746} (\bibinfo {year} {2024})}\BibitemShut {NoStop}%
\bibitem [{\citenamefont {Mitra}\ \emph {et~al.}(2024)\citenamefont {Mitra}, \citenamefont {Jim{\'e}nez-Gal{\'a}n}, \citenamefont {Aulich}, \citenamefont {Neuhaus}, \citenamefont {Silva}, \citenamefont {Pervak}, \citenamefont {Kling},\ and\ \citenamefont {Biswas}}]{Mitra2024}%
  \BibitemOpen
  \bibfield  {author} {\bibinfo {author} {\bibfnamefont {S.}~\bibnamefont {Mitra}}, \bibinfo {author} {\bibfnamefont {{\'A}.}~\bibnamefont {Jim{\'e}nez-Gal{\'a}n}}, \bibinfo {author} {\bibfnamefont {M.}~\bibnamefont {Aulich}}, \bibinfo {author} {\bibfnamefont {M.}~\bibnamefont {Neuhaus}}, \bibinfo {author} {\bibfnamefont {R.~E.~F.}\ \bibnamefont {Silva}}, \bibinfo {author} {\bibfnamefont {V.}~\bibnamefont {Pervak}}, \bibinfo {author} {\bibfnamefont {M.~F.}\ \bibnamefont {Kling}}, \ and\ \bibinfo {author} {\bibfnamefont {S.}~\bibnamefont {Biswas}},\ }\href {\doibase 10.1038/s41586-024-07244-z} {\bibfield  {journal} {\bibinfo  {journal} {Nature}\ }\textbf {\bibinfo {volume} {628}},\ \bibinfo {pages} {752} (\bibinfo {year} {2024})}\BibitemShut {NoStop}%
\bibitem [{\citenamefont {Zeng}\ \emph {et~al.}(2012)\citenamefont {Zeng}, \citenamefont {Dai}, \citenamefont {Yao}, \citenamefont {Xiao},\ and\ \citenamefont {Cui}}]{Zeng2012}%
  \BibitemOpen
  \bibfield  {author} {\bibinfo {author} {\bibfnamefont {H.}~\bibnamefont {Zeng}}, \bibinfo {author} {\bibfnamefont {J.}~\bibnamefont {Dai}}, \bibinfo {author} {\bibfnamefont {W.}~\bibnamefont {Yao}}, \bibinfo {author} {\bibfnamefont {D.}~\bibnamefont {Xiao}}, \ and\ \bibinfo {author} {\bibfnamefont {X.}~\bibnamefont {Cui}},\ }\href {\doibase 10.1038/nnano.2012.95} {\bibfield  {journal} {\bibinfo  {journal} {Nat. Nanotechnol.}\ }\textbf {\bibinfo {volume} {7}},\ \bibinfo {pages} {490} (\bibinfo {year} {2012})}\BibitemShut {NoStop}%
\bibitem [{\citenamefont {Mak}\ \emph {et~al.}(2012)\citenamefont {Mak}, \citenamefont {He}, \citenamefont {Shan},\ and\ \citenamefont {Heinz}}]{Mak2012}%
  \BibitemOpen
  \bibfield  {author} {\bibinfo {author} {\bibfnamefont {K.~F.}\ \bibnamefont {Mak}}, \bibinfo {author} {\bibfnamefont {K.}~\bibnamefont {He}}, \bibinfo {author} {\bibfnamefont {J.}~\bibnamefont {Shan}}, \ and\ \bibinfo {author} {\bibfnamefont {T.~F.}\ \bibnamefont {Heinz}},\ }\href {\doibase 10.1038/nnano.2012.96} {\bibfield  {journal} {\bibinfo  {journal} {Nat. Nanotechnol.}\ }\textbf {\bibinfo {volume} {7}},\ \bibinfo {pages} {494} (\bibinfo {year} {2012})}\BibitemShut {NoStop}%
\bibitem [{\citenamefont {Hsu}\ \emph {et~al.}(2015)\citenamefont {Hsu}, \citenamefont {Chen}, \citenamefont {Chen}, \citenamefont {Liu}, \citenamefont {Hou}, \citenamefont {Li},\ and\ \citenamefont {Chang}}]{Hsu2015}%
  \BibitemOpen
  \bibfield  {author} {\bibinfo {author} {\bibfnamefont {W.-T.}\ \bibnamefont {Hsu}}, \bibinfo {author} {\bibfnamefont {Y.-L.}\ \bibnamefont {Chen}}, \bibinfo {author} {\bibfnamefont {C.-H.}\ \bibnamefont {Chen}}, \bibinfo {author} {\bibfnamefont {P.-S.}\ \bibnamefont {Liu}}, \bibinfo {author} {\bibfnamefont {T.-H.}\ \bibnamefont {Hou}}, \bibinfo {author} {\bibfnamefont {L.-J.}\ \bibnamefont {Li}}, \ and\ \bibinfo {author} {\bibfnamefont {W.-H.}\ \bibnamefont {Chang}},\ }\href {https://doi.org/10.1038/ncomms9963} {\bibfield  {journal} {\bibinfo  {journal} {Nat. Commun.}\ }\textbf {\bibinfo {volume} {6}},\ \bibinfo {pages} {8963} (\bibinfo {year} {2015})}\BibitemShut {NoStop}%
\bibitem [{\citenamefont {Wang}\ \emph {et~al.}(2013)\citenamefont {Wang}, \citenamefont {Ge}, \citenamefont {Li}, \citenamefont {Qiu}, \citenamefont {Ji}, \citenamefont {Feng},\ and\ \citenamefont {Sun}}]{Wang2013}%
  \BibitemOpen
  \bibfield  {author} {\bibinfo {author} {\bibfnamefont {Q.}~\bibnamefont {Wang}}, \bibinfo {author} {\bibfnamefont {S.}~\bibnamefont {Ge}}, \bibinfo {author} {\bibfnamefont {X.}~\bibnamefont {Li}}, \bibinfo {author} {\bibfnamefont {J.}~\bibnamefont {Qiu}}, \bibinfo {author} {\bibfnamefont {Y.}~\bibnamefont {Ji}}, \bibinfo {author} {\bibfnamefont {J.}~\bibnamefont {Feng}}, \ and\ \bibinfo {author} {\bibfnamefont {D.}~\bibnamefont {Sun}},\ }\href {\doibase 10.1021/nn405419h} {\bibfield  {journal} {\bibinfo  {journal} {ACS Nano}\ }\textbf {\bibinfo {volume} {7}},\ \bibinfo {pages} {11087} (\bibinfo {year} {2013})}\BibitemShut {NoStop}%
\bibitem [{\citenamefont {Kumar}\ \emph {et~al.}(2014)\citenamefont {Kumar}, \citenamefont {He}, \citenamefont {He}, \citenamefont {Wang},\ and\ \citenamefont {Zhao}}]{Kumar2014}%
  \BibitemOpen
  \bibfield  {author} {\bibinfo {author} {\bibfnamefont {N.}~\bibnamefont {Kumar}}, \bibinfo {author} {\bibfnamefont {J.}~\bibnamefont {He}}, \bibinfo {author} {\bibfnamefont {D.}~\bibnamefont {He}}, \bibinfo {author} {\bibfnamefont {Y.}~\bibnamefont {Wang}}, \ and\ \bibinfo {author} {\bibfnamefont {H.}~\bibnamefont {Zhao}},\ }\href {\doibase 10.1039/c4nr03607g} {\bibfield  {journal} {\bibinfo  {journal} {Nanoscale}\ }\textbf {\bibinfo {volume} {6}},\ \bibinfo {pages} {12690} (\bibinfo {year} {2014})}\BibitemShut {NoStop}%
\bibitem [{\citenamefont {Mai}\ \emph {et~al.}(2014)\citenamefont {Mai}, \citenamefont {Semenov}, \citenamefont {Barrette}, \citenamefont {Yu}, \citenamefont {Jin}, \citenamefont {Cao}, \citenamefont {Kim},\ and\ \citenamefont {Gundogdu}}]{Mai2014}%
  \BibitemOpen
  \bibfield  {author} {\bibinfo {author} {\bibfnamefont {C.}~\bibnamefont {Mai}}, \bibinfo {author} {\bibfnamefont {Y.~G.}\ \bibnamefont {Semenov}}, \bibinfo {author} {\bibfnamefont {A.}~\bibnamefont {Barrette}}, \bibinfo {author} {\bibfnamefont {Y.}~\bibnamefont {Yu}}, \bibinfo {author} {\bibfnamefont {Z.}~\bibnamefont {Jin}}, \bibinfo {author} {\bibfnamefont {L.}~\bibnamefont {Cao}}, \bibinfo {author} {\bibfnamefont {K.~W.}\ \bibnamefont {Kim}}, \ and\ \bibinfo {author} {\bibfnamefont {K.}~\bibnamefont {Gundogdu}},\ }\href {\doibase 10.1103/physrevb.90.041414} {\bibfield  {journal} {\bibinfo  {journal} {Phys. Rev. B}\ }\textbf {\bibinfo {volume} {90}},\ \bibinfo {pages} {041414} (\bibinfo {year} {2014})}\BibitemShut {NoStop}%
\bibitem [{\citenamefont {Bertoni}\ \emph {et~al.}(2016)\citenamefont {Bertoni}, \citenamefont {Nicholson}, \citenamefont {Waldecker}, \citenamefont {Hübener}, \citenamefont {Monney}, \citenamefont {De~Giovannini}, \citenamefont {Puppin}, \citenamefont {Hoesch}, \citenamefont {Springate}, \citenamefont {Chapman}, \citenamefont {Cacho}, \citenamefont {Wolf}, \citenamefont {Rubio},\ and\ \citenamefont {Ernstorfer}}]{Bertoni2016}%
  \BibitemOpen
  \bibfield  {author} {\bibinfo {author} {\bibfnamefont {R.}~\bibnamefont {Bertoni}}, \bibinfo {author} {\bibfnamefont {C.}~\bibnamefont {Nicholson}}, \bibinfo {author} {\bibfnamefont {L.}~\bibnamefont {Waldecker}}, \bibinfo {author} {\bibfnamefont {H.}~\bibnamefont {Hübener}}, \bibinfo {author} {\bibfnamefont {C.}~\bibnamefont {Monney}}, \bibinfo {author} {\bibfnamefont {U.}~\bibnamefont {De~Giovannini}}, \bibinfo {author} {\bibfnamefont {M.}~\bibnamefont {Puppin}}, \bibinfo {author} {\bibfnamefont {M.}~\bibnamefont {Hoesch}}, \bibinfo {author} {\bibfnamefont {E.}~\bibnamefont {Springate}}, \bibinfo {author} {\bibfnamefont {R.}~\bibnamefont {Chapman}}, \bibinfo {author} {\bibfnamefont {C.}~\bibnamefont {Cacho}}, \bibinfo {author} {\bibfnamefont {M.}~\bibnamefont {Wolf}}, \bibinfo {author} {\bibfnamefont {A.}~\bibnamefont {Rubio}}, \ and\ \bibinfo {author} {\bibfnamefont {R.}~\bibnamefont {Ernstorfer}},\ }\href {\doibase 10.1103/physrevlett.117.277201} {\bibfield  {journal} {\bibinfo  {journal} {Phys. Rev.
  Lett.}\ }\textbf {\bibinfo {volume} {117}},\ \bibinfo {pages} {277201} (\bibinfo {year} {2016})}\BibitemShut {NoStop}%
\bibitem [{\citenamefont {Avetissian}\ \emph {et~al.}(2023)\citenamefont {Avetissian}, \citenamefont {Sedrakyan}, \citenamefont {Sedrakian},\ and\ \citenamefont {Mkrtchian}}]{Avetissian2023}%
  \BibitemOpen
  \bibfield  {author} {\bibinfo {author} {\bibfnamefont {H.~K.}\ \bibnamefont {Avetissian}}, \bibinfo {author} {\bibfnamefont {V.~A.}\ \bibnamefont {Sedrakyan}}, \bibinfo {author} {\bibfnamefont {K.~V.}\ \bibnamefont {Sedrakian}}, \ and\ \bibinfo {author} {\bibfnamefont {G.~F.}\ \bibnamefont {Mkrtchian}},\ }\href {\doibase 10.1103/physrevb.107.205403} {\bibfield  {journal} {\bibinfo  {journal} {Phys. Rev. B}\ }\textbf {\bibinfo {volume} {107}},\ \bibinfo {pages} {205403} (\bibinfo {year} {2023})}\BibitemShut {NoStop}%
\bibitem [{\citenamefont {Herrmann}\ \emph {et~al.}(2023)\citenamefont {Herrmann}, \citenamefont {Klimmer}, \citenamefont {Lettau}, \citenamefont {Monfared}, \citenamefont {Staude}, \citenamefont {Paradisanos}, \citenamefont {Peschel},\ and\ \citenamefont {Soavi}}]{Herrmann2023}%
  \BibitemOpen
  \bibfield  {author} {\bibinfo {author} {\bibfnamefont {P.}~\bibnamefont {Herrmann}}, \bibinfo {author} {\bibfnamefont {S.}~\bibnamefont {Klimmer}}, \bibinfo {author} {\bibfnamefont {T.}~\bibnamefont {Lettau}}, \bibinfo {author} {\bibfnamefont {M.}~\bibnamefont {Monfared}}, \bibinfo {author} {\bibfnamefont {I.}~\bibnamefont {Staude}}, \bibinfo {author} {\bibfnamefont {I.}~\bibnamefont {Paradisanos}}, \bibinfo {author} {\bibfnamefont {U.}~\bibnamefont {Peschel}}, \ and\ \bibinfo {author} {\bibfnamefont {G.}~\bibnamefont {Soavi}},\ }\href {https://doi.org/10.1002/smll.202301126} {\bibfield  {journal} {\bibinfo  {journal} {Small}\ }\textbf {\bibinfo {volume} {19}},\ \bibinfo {pages} {2301126} (\bibinfo {year} {2023})}\BibitemShut {NoStop}%
\bibitem [{\citenamefont {Wang}\ \emph {et~al.}(2018)\citenamefont {Wang}, \citenamefont {Molina-Sánchez}, \citenamefont {Altmann}, \citenamefont {Sangalli}, \citenamefont {De~Fazio}, \citenamefont {Soavi}, \citenamefont {Sassi}, \citenamefont {Bottegoni}, \citenamefont {Ciccacci}, \citenamefont {Finazzi}, \citenamefont {Wirtz}, \citenamefont {Ferrari}, \citenamefont {Marini}, \citenamefont {Cerullo},\ and\ \citenamefont {Dal~Conte}}]{Wang2018}%
  \BibitemOpen
  \bibfield  {author} {\bibinfo {author} {\bibfnamefont {Z.}~\bibnamefont {Wang}}, \bibinfo {author} {\bibfnamefont {A.}~\bibnamefont {Molina-Sánchez}}, \bibinfo {author} {\bibfnamefont {P.}~\bibnamefont {Altmann}}, \bibinfo {author} {\bibfnamefont {D.}~\bibnamefont {Sangalli}}, \bibinfo {author} {\bibfnamefont {D.}~\bibnamefont {De~Fazio}}, \bibinfo {author} {\bibfnamefont {G.}~\bibnamefont {Soavi}}, \bibinfo {author} {\bibfnamefont {U.}~\bibnamefont {Sassi}}, \bibinfo {author} {\bibfnamefont {F.}~\bibnamefont {Bottegoni}}, \bibinfo {author} {\bibfnamefont {F.}~\bibnamefont {Ciccacci}}, \bibinfo {author} {\bibfnamefont {M.}~\bibnamefont {Finazzi}}, \bibinfo {author} {\bibfnamefont {L.}~\bibnamefont {Wirtz}}, \bibinfo {author} {\bibfnamefont {A.~C.}\ \bibnamefont {Ferrari}}, \bibinfo {author} {\bibfnamefont {A.}~\bibnamefont {Marini}}, \bibinfo {author} {\bibfnamefont {G.}~\bibnamefont {Cerullo}}, \ and\ \bibinfo {author} {\bibfnamefont {S.}~\bibnamefont {Dal~Conte}},\ }\href {\doibase
  10.1021/acs.nanolett.8b02774} {\bibfield  {journal} {\bibinfo  {journal} {Nano Lett.}\ }\textbf {\bibinfo {volume} {18}},\ \bibinfo {pages} {6882} (\bibinfo {year} {2018})}\BibitemShut {NoStop}%
\bibitem [{\citenamefont {Neufeld}\ \emph {et~al.}(2023)\citenamefont {Neufeld}, \citenamefont {Tancogne-Dejean}, \citenamefont {Hübener}, \citenamefont {De~Giovannini},\ and\ \citenamefont {Rubio}}]{Neufeld2023}%
  \BibitemOpen
  \bibfield  {author} {\bibinfo {author} {\bibfnamefont {O.}~\bibnamefont {Neufeld}}, \bibinfo {author} {\bibfnamefont {N.}~\bibnamefont {Tancogne-Dejean}}, \bibinfo {author} {\bibfnamefont {H.}~\bibnamefont {Hübener}}, \bibinfo {author} {\bibfnamefont {U.}~\bibnamefont {De~Giovannini}}, \ and\ \bibinfo {author} {\bibfnamefont {A.}~\bibnamefont {Rubio}},\ }\href {\doibase 10.1103/physrevx.13.031011} {\bibfield  {journal} {\bibinfo  {journal} {Phys. Rev. X}\ }\textbf {\bibinfo {volume} {13}},\ \bibinfo {pages} {031011} (\bibinfo {year} {2023})}\BibitemShut {NoStop}%
\bibitem [{\citenamefont {Rana}\ and\ \citenamefont {Dixit}(2023)}]{Rana2023}%
  \BibitemOpen
  \bibfield  {author} {\bibinfo {author} {\bibfnamefont {N.}~\bibnamefont {Rana}}\ and\ \bibinfo {author} {\bibfnamefont {G.}~\bibnamefont {Dixit}},\ }\href {\doibase 10.1103/physrevapplied.19.034056} {\bibfield  {journal} {\bibinfo  {journal} {Phys. Rev. Applied}\ }\textbf {\bibinfo {volume} {19}},\ \bibinfo {pages} {034056} (\bibinfo {year} {2023})}\BibitemShut {NoStop}%
\bibitem [{\citenamefont {Wu}\ \emph {et~al.}(2016)\citenamefont {Wu}, \citenamefont {Chen}, \citenamefont {Camp}, \citenamefont {Schafer},\ and\ \citenamefont {Gaarde}}]{Wu2016}%
  \BibitemOpen
  \bibfield  {author} {\bibinfo {author} {\bibfnamefont {M.}~\bibnamefont {Wu}}, \bibinfo {author} {\bibfnamefont {S.}~\bibnamefont {Chen}}, \bibinfo {author} {\bibfnamefont {S.}~\bibnamefont {Camp}}, \bibinfo {author} {\bibfnamefont {K.~J.}\ \bibnamefont {Schafer}}, \ and\ \bibinfo {author} {\bibfnamefont {M.~B.}\ \bibnamefont {Gaarde}},\ }\href {\doibase 10.1088/0953-4075/49/6/062003} {\bibfield  {journal} {\bibinfo  {journal} {J. Phys. B: At. Mol. Opt. Phys.}\ }\textbf {\bibinfo {volume} {49}},\ \bibinfo {pages} {062003} (\bibinfo {year} {2016})}\BibitemShut {NoStop}%
\bibitem [{\citenamefont {R\o{}rstad}\ \emph {et~al.}(2017)\citenamefont {R\o{}rstad}, \citenamefont {B\ae{}kh\o{}j},\ and\ \citenamefont {Madsen}}]{Roerstad2017}%
  \BibitemOpen
  \bibfield  {author} {\bibinfo {author} {\bibfnamefont {J.~J.}\ \bibnamefont {R\o{}rstad}}, \bibinfo {author} {\bibfnamefont {J.~E.}\ \bibnamefont {B\ae{}kh\o{}j}}, \ and\ \bibinfo {author} {\bibfnamefont {L.~B.}\ \bibnamefont {Madsen}},\ }\href {\doibase 10.1103/physreva.96.013430} {\bibfield  {journal} {\bibinfo  {journal} {Phys. Rev. A}\ }\textbf {\bibinfo {volume} {96}},\ \bibinfo {pages} {013430} (\bibinfo {year} {2017})}\BibitemShut {NoStop}%
\bibitem [{\citenamefont {Fazzi}\ \emph {et~al.}(2013)\citenamefont {Fazzi}, \citenamefont {Scotognella}, \citenamefont {Milani}, \citenamefont {Brida}, \citenamefont {Manzoni}, \citenamefont {Cinquanta}, \citenamefont {Devetta}, \citenamefont {Ravagnan}, \citenamefont {Milani}, \citenamefont {Cataldo}, \citenamefont {Lüer}, \citenamefont {Wannemacher}, \citenamefont {Cabanillas-Gonzalez}, \citenamefont {Negro}, \citenamefont {Stagira},\ and\ \citenamefont {Vozzi}}]{Fazzi2013}%
  \BibitemOpen
  \bibfield  {author} {\bibinfo {author} {\bibfnamefont {D.}~\bibnamefont {Fazzi}}, \bibinfo {author} {\bibfnamefont {F.}~\bibnamefont {Scotognella}}, \bibinfo {author} {\bibfnamefont {A.}~\bibnamefont {Milani}}, \bibinfo {author} {\bibfnamefont {D.}~\bibnamefont {Brida}}, \bibinfo {author} {\bibfnamefont {C.}~\bibnamefont {Manzoni}}, \bibinfo {author} {\bibfnamefont {E.}~\bibnamefont {Cinquanta}}, \bibinfo {author} {\bibfnamefont {M.}~\bibnamefont {Devetta}}, \bibinfo {author} {\bibfnamefont {L.}~\bibnamefont {Ravagnan}}, \bibinfo {author} {\bibfnamefont {P.}~\bibnamefont {Milani}}, \bibinfo {author} {\bibfnamefont {F.}~\bibnamefont {Cataldo}}, \bibinfo {author} {\bibfnamefont {L.}~\bibnamefont {Lüer}}, \bibinfo {author} {\bibfnamefont {R.}~\bibnamefont {Wannemacher}}, \bibinfo {author} {\bibfnamefont {J.}~\bibnamefont {Cabanillas-Gonzalez}}, \bibinfo {author} {\bibfnamefont {M.}~\bibnamefont {Negro}}, \bibinfo {author} {\bibfnamefont {S.}~\bibnamefont {Stagira}}, \ and\ \bibinfo {author} {\bibfnamefont
  {C.}~\bibnamefont {Vozzi}},\ }\href {\doibase 10.1039/c3cp50508a} {\bibfield  {journal} {\bibinfo  {journal} {Phys. Chem. Chem. Phys.}\ }\textbf {\bibinfo {volume} {15}},\ \bibinfo {pages} {9384} (\bibinfo {year} {2013})}\BibitemShut {NoStop}%
\bibitem [{\citenamefont {Kfir}\ \emph {et~al.}(2014)\citenamefont {Kfir}, \citenamefont {Grychtol}, \citenamefont {Turgut}, \citenamefont {Knut}, \citenamefont {Zusin}, \citenamefont {Popmintchev}, \citenamefont {Popmintchev}, \citenamefont {Nembach}, \citenamefont {Shaw}, \citenamefont {Fleischer}, \citenamefont {Kapteyn}, \citenamefont {Murnane},\ and\ \citenamefont {Cohen}}]{Kfir2014}%
  \BibitemOpen
  \bibfield  {author} {\bibinfo {author} {\bibfnamefont {O.}~\bibnamefont {Kfir}}, \bibinfo {author} {\bibfnamefont {P.}~\bibnamefont {Grychtol}}, \bibinfo {author} {\bibfnamefont {E.}~\bibnamefont {Turgut}}, \bibinfo {author} {\bibfnamefont {R.}~\bibnamefont {Knut}}, \bibinfo {author} {\bibfnamefont {D.}~\bibnamefont {Zusin}}, \bibinfo {author} {\bibfnamefont {D.}~\bibnamefont {Popmintchev}}, \bibinfo {author} {\bibfnamefont {T.}~\bibnamefont {Popmintchev}}, \bibinfo {author} {\bibfnamefont {H.}~\bibnamefont {Nembach}}, \bibinfo {author} {\bibfnamefont {J.~M.}\ \bibnamefont {Shaw}}, \bibinfo {author} {\bibfnamefont {A.}~\bibnamefont {Fleischer}}, \bibinfo {author} {\bibfnamefont {H.}~\bibnamefont {Kapteyn}}, \bibinfo {author} {\bibfnamefont {M.}~\bibnamefont {Murnane}}, \ and\ \bibinfo {author} {\bibfnamefont {O.}~\bibnamefont {Cohen}},\ }\href {\doibase 10.1038/nphoton.2014.293} {\bibfield  {journal} {\bibinfo  {journal} {Nat. Photon.}\ }\textbf {\bibinfo {volume} {9}},\ \bibinfo {pages} {99} (\bibinfo
  {year} {2014})}\BibitemShut {NoStop}%
\bibitem [{\citenamefont {Zhu}\ \emph {et~al.}(2022)\citenamefont {Zhu}, \citenamefont {Lu},\ and\ \citenamefont {Lein}}]{Zhu2022}%
  \BibitemOpen
  \bibfield  {author} {\bibinfo {author} {\bibfnamefont {X.}~\bibnamefont {Zhu}}, \bibinfo {author} {\bibfnamefont {P.}~\bibnamefont {Lu}}, \ and\ \bibinfo {author} {\bibfnamefont {M.}~\bibnamefont {Lein}},\ }\href {\doibase 10.1103/physrevlett.128.030401} {\bibfield  {journal} {\bibinfo  {journal} {Phys. Rev. Lett.}\ }\textbf {\bibinfo {volume} {128}},\ \bibinfo {pages} {030401} (\bibinfo {year} {2022})}\BibitemShut {NoStop}%
\bibitem [{\citenamefont {Geondzhian}\ \emph {et~al.}(2022)\citenamefont {Geondzhian}, \citenamefont {Rubio},\ and\ \citenamefont {Altarelli}}]{Geondzhian2022}%
  \BibitemOpen
  \bibfield  {author} {\bibinfo {author} {\bibfnamefont {A.}~\bibnamefont {Geondzhian}}, \bibinfo {author} {\bibfnamefont {A.}~\bibnamefont {Rubio}}, \ and\ \bibinfo {author} {\bibfnamefont {M.}~\bibnamefont {Altarelli}},\ }\href {\doibase 10.1103/physrevb.106.115433} {\bibfield  {journal} {\bibinfo  {journal} {Phys. Rev. B}\ }\textbf {\bibinfo {volume} {106}},\ \bibinfo {pages} {115433} (\bibinfo {year} {2022})}\BibitemShut {NoStop}%
\bibitem [{\citenamefont {Feldman}\ \emph {et~al.}(2024)\citenamefont {Feldman}, \citenamefont {Even~Tzur},\ and\ \citenamefont {Cohen}}]{Feldman2024}%
  \BibitemOpen
  \bibfield  {author} {\bibinfo {author} {\bibfnamefont {M.}~\bibnamefont {Feldman}}, \bibinfo {author} {\bibfnamefont {M.}~\bibnamefont {Even~Tzur}}, \ and\ \bibinfo {author} {\bibfnamefont {O.}~\bibnamefont {Cohen}},\ }\href {\doibase 10.3390/photonics11100918} {\bibfield  {journal} {\bibinfo  {journal} {Photonics}\ }\textbf {\bibinfo {volume} {11}},\ \bibinfo {pages} {918} (\bibinfo {year} {2024})}\BibitemShut {NoStop}%
\bibitem [{\citenamefont {Siegrist}\ \emph {et~al.}(2019)\citenamefont {Siegrist}, \citenamefont {Gessner}, \citenamefont {Ossiander}, \citenamefont {Denker}, \citenamefont {Chang}, \citenamefont {Schröder}, \citenamefont {Guggenmos}, \citenamefont {Cui}, \citenamefont {Walowski}, \citenamefont {Martens}, \citenamefont {Dewhurst}, \citenamefont {Kleineberg}, \citenamefont {Münzenberg}, \citenamefont {Sharma},\ and\ \citenamefont {Schultze}}]{Siegrist2019}%
  \BibitemOpen
  \bibfield  {author} {\bibinfo {author} {\bibfnamefont {F.}~\bibnamefont {Siegrist}}, \bibinfo {author} {\bibfnamefont {J.~A.}\ \bibnamefont {Gessner}}, \bibinfo {author} {\bibfnamefont {M.}~\bibnamefont {Ossiander}}, \bibinfo {author} {\bibfnamefont {C.}~\bibnamefont {Denker}}, \bibinfo {author} {\bibfnamefont {Y.-P.}\ \bibnamefont {Chang}}, \bibinfo {author} {\bibfnamefont {M.~C.}\ \bibnamefont {Schröder}}, \bibinfo {author} {\bibfnamefont {A.}~\bibnamefont {Guggenmos}}, \bibinfo {author} {\bibfnamefont {Y.}~\bibnamefont {Cui}}, \bibinfo {author} {\bibfnamefont {J.}~\bibnamefont {Walowski}}, \bibinfo {author} {\bibfnamefont {U.}~\bibnamefont {Martens}}, \bibinfo {author} {\bibfnamefont {J.~K.}\ \bibnamefont {Dewhurst}}, \bibinfo {author} {\bibfnamefont {U.}~\bibnamefont {Kleineberg}}, \bibinfo {author} {\bibfnamefont {M.}~\bibnamefont {Münzenberg}}, \bibinfo {author} {\bibfnamefont {S.}~\bibnamefont {Sharma}}, \ and\ \bibinfo {author} {\bibfnamefont {M.}~\bibnamefont {Schultze}},\ }\href {\doibase
  10.1038/s41586-019-1333-x} {\bibfield  {journal} {\bibinfo  {journal} {Nature}\ }\textbf {\bibinfo {volume} {571}},\ \bibinfo {pages} {240} (\bibinfo {year} {2019})}\BibitemShut {NoStop}%
\bibitem [{\citenamefont {Han}\ \emph {et~al.}(2023)\citenamefont {Han}, \citenamefont {Ji}, \citenamefont {Ueda},\ and\ \citenamefont {Wörner}}]{Han2023}%
  \BibitemOpen
  \bibfield  {author} {\bibinfo {author} {\bibfnamefont {M.}~\bibnamefont {Han}}, \bibinfo {author} {\bibfnamefont {J.-B.}\ \bibnamefont {Ji}}, \bibinfo {author} {\bibfnamefont {K.}~\bibnamefont {Ueda}}, \ and\ \bibinfo {author} {\bibfnamefont {H.~J.}\ \bibnamefont {Wörner}},\ }\href {\doibase 10.1364/optica.492741} {\bibfield  {journal} {\bibinfo  {journal} {Optica}\ }\textbf {\bibinfo {volume} {10}},\ \bibinfo {pages} {1044} (\bibinfo {year} {2023})}\BibitemShut {NoStop}%
\bibitem [{\citenamefont {G{\'e}neaux}\ \emph {et~al.}(2024)\citenamefont {G{\'e}neaux}, \citenamefont {Chang}, \citenamefont {Guggenmos}, \citenamefont {Delaunay}, \citenamefont {L{\'e}gar{\'e}}, \citenamefont {L{\'e}gar{\'e}}, \citenamefont {L\"{u}ning}, \citenamefont {Parpiiev}, \citenamefont {Molesky}, \citenamefont {de~Roulet}, \citenamefont {Zuerch}, \citenamefont {Sharma}, \citenamefont {Schultze},\ and\ \citenamefont {Leone}}]{Geneaux2024}%
  \BibitemOpen
  \bibfield  {author} {\bibinfo {author} {\bibfnamefont {R.}~\bibnamefont {G{\'e}neaux}}, \bibinfo {author} {\bibfnamefont {H.-T.}\ \bibnamefont {Chang}}, \bibinfo {author} {\bibfnamefont {A.}~\bibnamefont {Guggenmos}}, \bibinfo {author} {\bibfnamefont {R.}~\bibnamefont {Delaunay}}, \bibinfo {author} {\bibfnamefont {F.}~\bibnamefont {L{\'e}gar{\'e}}}, \bibinfo {author} {\bibfnamefont {K.}~\bibnamefont {L{\'e}gar{\'e}}}, \bibinfo {author} {\bibfnamefont {J.}~\bibnamefont {L\"{u}ning}}, \bibinfo {author} {\bibfnamefont {T.}~\bibnamefont {Parpiiev}}, \bibinfo {author} {\bibfnamefont {I.~J.}\ \bibnamefont {Molesky}}, \bibinfo {author} {\bibfnamefont {B.~R.}\ \bibnamefont {de~Roulet}}, \bibinfo {author} {\bibfnamefont {M.~W.}\ \bibnamefont {Zuerch}}, \bibinfo {author} {\bibfnamefont {S.}~\bibnamefont {Sharma}}, \bibinfo {author} {\bibfnamefont {M.}~\bibnamefont {Schultze}}, \ and\ \bibinfo {author} {\bibfnamefont {S.~R.}\ \bibnamefont {Leone}},\ }\href {\doibase 10.1103/physrevlett.133.106902} {\bibfield
  {journal} {\bibinfo  {journal} {Phys. Rev. Lett.}\ }\textbf {\bibinfo {volume} {133}},\ \bibinfo {pages} {106902} (\bibinfo {year} {2024})}\BibitemShut {NoStop}%
\bibitem [{Sup()}]{Supp}%
  \BibitemOpen
  \href@noop {} {}\bibinfo {note} {See Supplemental Material for a detailed derivation of our scheme and details of the simulations, which includes Refs.\cite{ Marques2003, Castro2006, Andrade2015, TancogneDejean2020a, Perdew1996, Schlipf2015, Sato2021, TancogneDejean2020, Gaarde2011, Yamada2023, Cavaletto2024, Sato2022}.}\BibitemShut {Stop}%
\bibitem [{\citenamefont {Marques}(2003)}]{Marques2003}%
  \BibitemOpen
  \bibfield  {author} {\bibinfo {author} {\bibfnamefont {M.}~\bibnamefont {Marques}},\ }\href {\doibase 10.1016/s0010-4655(02)00686-0} {\bibfield  {journal} {\bibinfo  {journal} {Comput. Phys. Commun.}\ }\textbf {\bibinfo {volume} {151}},\ \bibinfo {pages} {60} (\bibinfo {year} {2003})}\BibitemShut {NoStop}%
\bibitem [{\citenamefont {Castro}\ \emph {et~al.}(2006)\citenamefont {Castro}, \citenamefont {Appel}, \citenamefont {Oliveira}, \citenamefont {Rozzi}, \citenamefont {Andrade}, \citenamefont {Lorenzen}, \citenamefont {Marques}, \citenamefont {Gross},\ and\ \citenamefont {Rubio}}]{Castro2006}%
  \BibitemOpen
  \bibfield  {author} {\bibinfo {author} {\bibfnamefont {A.}~\bibnamefont {Castro}}, \bibinfo {author} {\bibfnamefont {H.}~\bibnamefont {Appel}}, \bibinfo {author} {\bibfnamefont {M.}~\bibnamefont {Oliveira}}, \bibinfo {author} {\bibfnamefont {C.~A.}\ \bibnamefont {Rozzi}}, \bibinfo {author} {\bibfnamefont {X.}~\bibnamefont {Andrade}}, \bibinfo {author} {\bibfnamefont {F.}~\bibnamefont {Lorenzen}}, \bibinfo {author} {\bibfnamefont {M.~A.~L.}\ \bibnamefont {Marques}}, \bibinfo {author} {\bibfnamefont {E.~K.~U.}\ \bibnamefont {Gross}}, \ and\ \bibinfo {author} {\bibfnamefont {A.}~\bibnamefont {Rubio}},\ }\href {\doibase 10.1002/pssb.200642067} {\bibfield  {journal} {\bibinfo  {journal} {Phys. Status Solidi B}\ }\textbf {\bibinfo {volume} {243}},\ \bibinfo {pages} {2465} (\bibinfo {year} {2006})}\BibitemShut {NoStop}%
\bibitem [{\citenamefont {Andrade}\ \emph {et~al.}(2015)\citenamefont {Andrade}, \citenamefont {Strubbe}, \citenamefont {De~Giovannini}, \citenamefont {Larsen}, \citenamefont {Oliveira}, \citenamefont {Alberdi-Rodriguez}, \citenamefont {Varas}, \citenamefont {Theophilou}, \citenamefont {Helbig}, \citenamefont {Verstraete}, \citenamefont {Stella}, \citenamefont {Nogueira}, \citenamefont {Aspuru-Guzik}, \citenamefont {Castro}, \citenamefont {Marques},\ and\ \citenamefont {Rubio}}]{Andrade2015}%
  \BibitemOpen
  \bibfield  {author} {\bibinfo {author} {\bibfnamefont {X.}~\bibnamefont {Andrade}}, \bibinfo {author} {\bibfnamefont {D.}~\bibnamefont {Strubbe}}, \bibinfo {author} {\bibfnamefont {U.}~\bibnamefont {De~Giovannini}}, \bibinfo {author} {\bibfnamefont {A.~H.}\ \bibnamefont {Larsen}}, \bibinfo {author} {\bibfnamefont {M.~J.~T.}\ \bibnamefont {Oliveira}}, \bibinfo {author} {\bibfnamefont {J.}~\bibnamefont {Alberdi-Rodriguez}}, \bibinfo {author} {\bibfnamefont {A.}~\bibnamefont {Varas}}, \bibinfo {author} {\bibfnamefont {I.}~\bibnamefont {Theophilou}}, \bibinfo {author} {\bibfnamefont {N.}~\bibnamefont {Helbig}}, \bibinfo {author} {\bibfnamefont {M.~J.}\ \bibnamefont {Verstraete}}, \bibinfo {author} {\bibfnamefont {L.}~\bibnamefont {Stella}}, \bibinfo {author} {\bibfnamefont {F.}~\bibnamefont {Nogueira}}, \bibinfo {author} {\bibfnamefont {A.}~\bibnamefont {Aspuru-Guzik}}, \bibinfo {author} {\bibfnamefont {A.}~\bibnamefont {Castro}}, \bibinfo {author} {\bibfnamefont {M.~A.~L.}\ \bibnamefont {Marques}}, \ and\
  \bibinfo {author} {\bibfnamefont {A.}~\bibnamefont {Rubio}},\ }\href {\doibase 10.1039/c5cp00351b} {\bibfield  {journal} {\bibinfo  {journal} {Phys. Chem. Chem. Phys.}\ }\textbf {\bibinfo {volume} {17}},\ \bibinfo {pages} {31371} (\bibinfo {year} {2015})}\BibitemShut {NoStop}%
\bibitem [{\citenamefont {Tancogne-Dejean}\ \emph {et~al.}(2020{\natexlab{a}})\citenamefont {Tancogne-Dejean}, \citenamefont {Oliveira}, \citenamefont {Andrade}, \citenamefont {Appel}, \citenamefont {Borca}, \citenamefont {Le~Breton}, \citenamefont {Buchholz}, \citenamefont {Castro}, \citenamefont {Corni}, \citenamefont {Correa}, \citenamefont {De~Giovannini}, \citenamefont {Delgado}, \citenamefont {Eich}, \citenamefont {Flick}, \citenamefont {Gil}, \citenamefont {Gomez}, \citenamefont {Helbig}, \citenamefont {Hübener}, \citenamefont {Jestädt}, \citenamefont {Jornet-Somoza}, \citenamefont {Larsen}, \citenamefont {Lebedeva}, \citenamefont {Lüders}, \citenamefont {Marques}, \citenamefont {Ohlmann}, \citenamefont {Pipolo}, \citenamefont {Rampp}, \citenamefont {Rozzi}, \citenamefont {Strubbe}, \citenamefont {Sato}, \citenamefont {Schäfer}, \citenamefont {Theophilou}, \citenamefont {Welden},\ and\ \citenamefont {Rubio}}]{TancogneDejean2020a}%
  \BibitemOpen
  \bibfield  {author} {\bibinfo {author} {\bibfnamefont {N.}~\bibnamefont {Tancogne-Dejean}}, \bibinfo {author} {\bibfnamefont {M.~J.~T.}\ \bibnamefont {Oliveira}}, \bibinfo {author} {\bibfnamefont {X.}~\bibnamefont {Andrade}}, \bibinfo {author} {\bibfnamefont {H.}~\bibnamefont {Appel}}, \bibinfo {author} {\bibfnamefont {C.~H.}\ \bibnamefont {Borca}}, \bibinfo {author} {\bibfnamefont {G.}~\bibnamefont {Le~Breton}}, \bibinfo {author} {\bibfnamefont {F.}~\bibnamefont {Buchholz}}, \bibinfo {author} {\bibfnamefont {A.}~\bibnamefont {Castro}}, \bibinfo {author} {\bibfnamefont {S.}~\bibnamefont {Corni}}, \bibinfo {author} {\bibfnamefont {A.~A.}\ \bibnamefont {Correa}}, \bibinfo {author} {\bibfnamefont {U.}~\bibnamefont {De~Giovannini}}, \bibinfo {author} {\bibfnamefont {A.}~\bibnamefont {Delgado}}, \bibinfo {author} {\bibfnamefont {F.~G.}\ \bibnamefont {Eich}}, \bibinfo {author} {\bibfnamefont {J.}~\bibnamefont {Flick}}, \bibinfo {author} {\bibfnamefont {G.}~\bibnamefont {Gil}}, \bibinfo {author} {\bibfnamefont
  {A.}~\bibnamefont {Gomez}}, \bibinfo {author} {\bibfnamefont {N.}~\bibnamefont {Helbig}}, \bibinfo {author} {\bibfnamefont {H.}~\bibnamefont {Hübener}}, \bibinfo {author} {\bibfnamefont {R.}~\bibnamefont {Jestädt}}, \bibinfo {author} {\bibfnamefont {J.}~\bibnamefont {Jornet-Somoza}}, \bibinfo {author} {\bibfnamefont {A.~H.}\ \bibnamefont {Larsen}}, \bibinfo {author} {\bibfnamefont {I.~V.}\ \bibnamefont {Lebedeva}}, \bibinfo {author} {\bibfnamefont {M.}~\bibnamefont {Lüders}}, \bibinfo {author} {\bibfnamefont {M.~A.~L.}\ \bibnamefont {Marques}}, \bibinfo {author} {\bibfnamefont {S.~T.}\ \bibnamefont {Ohlmann}}, \bibinfo {author} {\bibfnamefont {S.}~\bibnamefont {Pipolo}}, \bibinfo {author} {\bibfnamefont {M.}~\bibnamefont {Rampp}}, \bibinfo {author} {\bibfnamefont {C.~A.}\ \bibnamefont {Rozzi}}, \bibinfo {author} {\bibfnamefont {D.~A.}\ \bibnamefont {Strubbe}}, \bibinfo {author} {\bibfnamefont {S.~A.}\ \bibnamefont {Sato}}, \bibinfo {author} {\bibfnamefont {C.}~\bibnamefont {Schäfer}}, \bibinfo {author}
  {\bibfnamefont {I.}~\bibnamefont {Theophilou}}, \bibinfo {author} {\bibfnamefont {A.}~\bibnamefont {Welden}}, \ and\ \bibinfo {author} {\bibfnamefont {A.}~\bibnamefont {Rubio}},\ }\href {https://doi.org/10.1063/1.5142502} {\bibfield  {journal} {\bibinfo  {journal} {J. Chem. Phys.}\ }\textbf {\bibinfo {volume} {152}} (\bibinfo {year} {2020}{\natexlab{a}})}\BibitemShut {NoStop}%
\bibitem [{\citenamefont {Berkelbach}\ \emph {et~al.}(2015)\citenamefont {Berkelbach}, \citenamefont {Hybertsen},\ and\ \citenamefont {Reichman}}]{Berkelbach2015}%
  \BibitemOpen
  \bibfield  {author} {\bibinfo {author} {\bibfnamefont {T.~C.}\ \bibnamefont {Berkelbach}}, \bibinfo {author} {\bibfnamefont {M.~S.}\ \bibnamefont {Hybertsen}}, \ and\ \bibinfo {author} {\bibfnamefont {D.~R.}\ \bibnamefont {Reichman}},\ }\href {\doibase 10.1103/physrevb.92.085413} {\bibfield  {journal} {\bibinfo  {journal} {Phys. Rev. B}\ }\textbf {\bibinfo {volume} {92}},\ \bibinfo {pages} {085413} (\bibinfo {year} {2015})}\BibitemShut {NoStop}%
\bibitem [{\citenamefont {Li}\ \emph {et~al.}(2019)\citenamefont {Li}, \citenamefont {Lan}, \citenamefont {Zhu}, \citenamefont {Huang}, \citenamefont {Zhang}, \citenamefont {Lein},\ and\ \citenamefont {Lu}}]{Li2019}%
  \BibitemOpen
  \bibfield  {author} {\bibinfo {author} {\bibfnamefont {L.}~\bibnamefont {Li}}, \bibinfo {author} {\bibfnamefont {P.}~\bibnamefont {Lan}}, \bibinfo {author} {\bibfnamefont {X.}~\bibnamefont {Zhu}}, \bibinfo {author} {\bibfnamefont {T.}~\bibnamefont {Huang}}, \bibinfo {author} {\bibfnamefont {Q.}~\bibnamefont {Zhang}}, \bibinfo {author} {\bibfnamefont {M.}~\bibnamefont {Lein}}, \ and\ \bibinfo {author} {\bibfnamefont {P.}~\bibnamefont {Lu}},\ }\href {\doibase 10.1103/physrevlett.122.193901} {\bibfield  {journal} {\bibinfo  {journal} {Phys. Rev. Lett.}\ }\textbf {\bibinfo {volume} {122}},\ \bibinfo {pages} {193901} (\bibinfo {year} {2019})}\BibitemShut {NoStop}%
\bibitem [{\citenamefont {Li}\ \emph {et~al.}(2023)\citenamefont {Li}, \citenamefont {Lan}, \citenamefont {Zhu},\ and\ \citenamefont {Lu}}]{Li2023}%
  \BibitemOpen
  \bibfield  {author} {\bibinfo {author} {\bibfnamefont {L.}~\bibnamefont {Li}}, \bibinfo {author} {\bibfnamefont {P.}~\bibnamefont {Lan}}, \bibinfo {author} {\bibfnamefont {X.}~\bibnamefont {Zhu}}, \ and\ \bibinfo {author} {\bibfnamefont {P.}~\bibnamefont {Lu}},\ }\href {\doibase 10.1088/1361-6633/acf144} {\bibfield  {journal} {\bibinfo  {journal} {Rep. Prog. Phys.}\ }\textbf {\bibinfo {volume} {86}},\ \bibinfo {pages} {116401} (\bibinfo {year} {2023})}\BibitemShut {NoStop}%
\bibitem [{\citenamefont {Lively}\ \emph {et~al.}(2024)\citenamefont {Lively}, \citenamefont {Sato}, \citenamefont {Albareda}, \citenamefont {Rubio},\ and\ \citenamefont {Kelly}}]{Lively2024}%
  \BibitemOpen
  \bibfield  {author} {\bibinfo {author} {\bibfnamefont {K.}~\bibnamefont {Lively}}, \bibinfo {author} {\bibfnamefont {S.~A.}\ \bibnamefont {Sato}}, \bibinfo {author} {\bibfnamefont {G.}~\bibnamefont {Albareda}}, \bibinfo {author} {\bibfnamefont {A.}~\bibnamefont {Rubio}}, \ and\ \bibinfo {author} {\bibfnamefont {A.}~\bibnamefont {Kelly}},\ }\href {\doibase 10.1103/physrevresearch.6.013069} {\bibfield  {journal} {\bibinfo  {journal} {Phys. Rev. Res.}\ }\textbf {\bibinfo {volume} {6}},\ \bibinfo {pages} {013069} (\bibinfo {year} {2024})}\BibitemShut {NoStop}%
\bibitem [{\citenamefont {Perdew}\ \emph {et~al.}(1996)\citenamefont {Perdew}, \citenamefont {Burke},\ and\ \citenamefont {Ernzerhof}}]{Perdew1996}%
  \BibitemOpen
  \bibfield  {author} {\bibinfo {author} {\bibfnamefont {J.~P.}\ \bibnamefont {Perdew}}, \bibinfo {author} {\bibfnamefont {K.}~\bibnamefont {Burke}}, \ and\ \bibinfo {author} {\bibfnamefont {M.}~\bibnamefont {Ernzerhof}},\ }\href {\doibase 10.1103/physrevlett.77.3865} {\bibfield  {journal} {\bibinfo  {journal} {Phys. Rev. Lett.}\ }\textbf {\bibinfo {volume} {77}},\ \bibinfo {pages} {3865} (\bibinfo {year} {1996})}\BibitemShut {NoStop}%
\bibitem [{\citenamefont {Schlipf}\ and\ \citenamefont {Gygi}(2015)}]{Schlipf2015}%
  \BibitemOpen
  \bibfield  {author} {\bibinfo {author} {\bibfnamefont {M.}~\bibnamefont {Schlipf}}\ and\ \bibinfo {author} {\bibfnamefont {F.}~\bibnamefont {Gygi}},\ }\href {\doibase 10.1016/j.cpc.2015.05.011} {\bibfield  {journal} {\bibinfo  {journal} {Comput. Phys. Commun.}\ }\textbf {\bibinfo {volume} {196}},\ \bibinfo {pages} {36} (\bibinfo {year} {2015})}\BibitemShut {NoStop}%
\bibitem [{\citenamefont {Sato}(2021)}]{Sato2021}%
  \BibitemOpen
  \bibfield  {author} {\bibinfo {author} {\bibfnamefont {S.~A.}\ \bibnamefont {Sato}},\ }\href {\doibase 10.1016/j.commatsci.2020.110274} {\bibfield  {journal} {\bibinfo  {journal} {Comput. Mater. Sci.}\ }\textbf {\bibinfo {volume} {194}},\ \bibinfo {pages} {110274} (\bibinfo {year} {2021})}\BibitemShut {NoStop}%
\bibitem [{\citenamefont {Tancogne-Dejean}\ \emph {et~al.}(2020{\natexlab{b}})\citenamefont {Tancogne-Dejean}, \citenamefont {Sentef},\ and\ \citenamefont {Rubio}}]{TancogneDejean2020}%
  \BibitemOpen
  \bibfield  {author} {\bibinfo {author} {\bibfnamefont {N.}~\bibnamefont {Tancogne-Dejean}}, \bibinfo {author} {\bibfnamefont {M.~A.}\ \bibnamefont {Sentef}}, \ and\ \bibinfo {author} {\bibfnamefont {A.}~\bibnamefont {Rubio}},\ }\href {\doibase 10.1103/physrevb.102.115106} {\bibfield  {journal} {\bibinfo  {journal} {Phys. Rev. B}\ }\textbf {\bibinfo {volume} {102}},\ \bibinfo {pages} {115106} (\bibinfo {year} {2020}{\natexlab{b}})}\BibitemShut {NoStop}%
\bibitem [{\citenamefont {Gaarde}\ \emph {et~al.}(2011)\citenamefont {Gaarde}, \citenamefont {Buth}, \citenamefont {Tate},\ and\ \citenamefont {Schafer}}]{Gaarde2011}%
  \BibitemOpen
  \bibfield  {author} {\bibinfo {author} {\bibfnamefont {M.~B.}\ \bibnamefont {Gaarde}}, \bibinfo {author} {\bibfnamefont {C.}~\bibnamefont {Buth}}, \bibinfo {author} {\bibfnamefont {J.~L.}\ \bibnamefont {Tate}}, \ and\ \bibinfo {author} {\bibfnamefont {K.~J.}\ \bibnamefont {Schafer}},\ }\href {\doibase 10.1103/physreva.83.013419} {\bibfield  {journal} {\bibinfo  {journal} {Phys. Rev. A}\ }\textbf {\bibinfo {volume} {83}},\ \bibinfo {pages} {013419} (\bibinfo {year} {2011})}\BibitemShut {NoStop}%
\bibitem [{\citenamefont {Yamada}\ \emph {et~al.}(2023)\citenamefont {Yamada}, \citenamefont {Yabana},\ and\ \citenamefont {Otobe}}]{Yamada2023}%
  \BibitemOpen
  \bibfield  {author} {\bibinfo {author} {\bibfnamefont {S.}~\bibnamefont {Yamada}}, \bibinfo {author} {\bibfnamefont {K.}~\bibnamefont {Yabana}}, \ and\ \bibinfo {author} {\bibfnamefont {T.}~\bibnamefont {Otobe}},\ }\href {\doibase 10.1103/physrevb.108.035404} {\bibfield  {journal} {\bibinfo  {journal} {Phys. Rev. B}\ }\textbf {\bibinfo {volume} {108}},\ \bibinfo {pages} {035404} (\bibinfo {year} {2023})}\BibitemShut {NoStop}%
\bibitem [{\citenamefont {Cavaletto}\ and\ \citenamefont {Madsen}(2024)}]{Cavaletto2024}%
  \BibitemOpen
  \bibfield  {author} {\bibinfo {author} {\bibfnamefont {S.~M.}\ \bibnamefont {Cavaletto}}\ and\ \bibinfo {author} {\bibfnamefont {L.~B.}\ \bibnamefont {Madsen}},\ }\href {\doibase 10.1103/physreva.110.053111} {\bibfield  {journal} {\bibinfo  {journal} {Phys. Rev. A}\ }\textbf {\bibinfo {volume} {110}},\ \bibinfo {pages} {053111} (\bibinfo {year} {2024})}\BibitemShut {NoStop}%
\bibitem [{\citenamefont {Sato}(2022)}]{Sato2022}%
  \BibitemOpen
  \bibfield  {author} {\bibinfo {author} {\bibfnamefont {S.~A.}\ \bibnamefont {Sato}},\ }\href {\doibase 10.1088/2516-1075/ac52df} {\bibfield  {journal} {\bibinfo  {journal} {Electron. Struct.}\ }\textbf {\bibinfo {volume} {4}},\ \bibinfo {pages} {014007} (\bibinfo {year} {2022})}\BibitemShut {NoStop}%
\end{thebibliography}%
\end{document}